\documentclass[numberedappendix]{emulateapj}

\begin{document}

\title{Energetic Impact of Jet Inflated Cocoons in Relaxed Galaxy Clusters}
\author{John C. Vernaleo and Christopher S. Reynolds}
\affil{Department of Astronomy, University of Maryland, College Park,
  MD 20742-2421}
\email{vernaleo@astro.umd.edu, chris@astro.umd.edu}

\begin{abstract}
Jets from active galactic nuclei (AGN) in the cores of galaxy clusters
have the potential to be a major contributor to the energy budget of
the intracluster medium (ICM).  To study the dependence of the
interaction between the AGN jets and the ICM on the parameters of the
jets themselves, we present a parameter survey of two-dimensional
(axisymmetric) ideal hydrodynamic models of back-to-back jets injected
into a cluster atmosphere (with varying Mach numbers and kinetic
luminosities).  We follow the passive evolution of the resulting
structures for several times longer than the active lifetime of the
jet.  The simulations fall into roughly two classes, cocoon-bounded
and non-cocoon bounded sources.  We suggest a correspondence between
these two classes and the Faranoff-Riley types.  We find that the
cocoon-bounded sources inject significantly more entropy into the core
regions of the ICM atmosphere, even though the efficiency with which
energy is thermalized is independent of the morphological class.  In
all cases, a large fraction (50--80\%) of the energy injected by the
jet ends up as gravitational potential energy due to the expansion of
the atmosphere.
\end{abstract}


\section{Introduction}
\label{sec:intro}

Jets from powerful Active Galactic Nuclei (AGN) play a major role in
shaping their environment on both the large and small scales.  In many
clusters of galaxies, the intracluster medium (ICM) bears the marks of
repeated episodes of AGN jet activity: e.g., bubbles
\citep{2000MNRAS.318L..65F, 2000ApJ...534L.135M, 2001ApJ...558L..15B,
2002ApJ...579..560Y}, ghost bubbles \citep{2001ApJ...562L.149M,
2002ApJ...569L..79H, 2004ApJ...606..185C, 2000MNRAS.318L..65F},
ripples \citep{2003MNRAS.344L..43F,2005MNRAS.360L..20F}, shells
\citep{2000MNRAS.318L..65F}, and filaments.  Besides affecting the
gross morphology of the ICM, AGN have the potential to dramatically
alter the energy and entropy budgets of the ICM.  This is especially
of interest in relation to the cooling flow problem of galaxy
clusters~\citep{1977ApJ...215..723C, 1977MNRAS.180..479F,
1994ARA&A..32..277F}.  In this paper, we attempt to determine the
relationship between the jet parameters (Mach number and kinetic
luminosity) and the resulting energetics and structures in the ICM.

X-ray observations show that the ICM radiates strongly in the X-ray
wavelengths due to thermal bremsstrahlung radiation.  In the central
regions of rich, relaxed clusters, the ICM has a cooling time shorter
than the Hubble time (sometimes as low as a few $\times 10^8$ years).
With such short cooling times, we would expect to see large quantities
of cool gas and/or star formation in the center of rich clusters.
Equivalently, the cD galaxy at the center of the cluster would be
accreting cooled matter from the cooling ICM.  The density-squared
dependence of the emission ensures that, if this simple picture were
correct, the cooling would increase and eventually reach a cooling
catastrophe in a finite time.   

Observations strongly suggest that large quantities of cooled gas are
not being deposited in the central galaxy.  The observed star
formation rates in the central cD galaxies are not sufficient to match
the inferred cooling
rates~\citep{2004ApJ...612..131O,2005ApJ...635L...9H}.  Furthermore,
the mass function of galaxies shows a high-mass truncation that argues
against the continued ICM accretion growth of cD galaxies and requires
a feedback process significantly more efficient than star
formation (see~\citet{2003ApJ...599...38B}).  More directly,
high-resolution X-ray spectroscopy with the Reflection Grating
Spectrometer (RGS) on {\it XMM-Newton} shows that the ICM cools from
the virial temperature to approximately one third of the virial
temperature but reveals an absence of plasma below this temperature.
The apparent contradiction between the absence of cool gas and the
short cooling times of the ICM core is the cooling flow
problem~\citep{1994ARA&A..32..277F}.

These observational results suggest that some form of ICM heating is
required to offset the radiative cooling.  Intermittent activity by a
central AGN remains an attractive solution.  A large amount of
theoretical work has been performed on the possible effects of AGN on
cooling cluster cores.  With ever increasing computer power, most of
the recent work has focused on hydrodynamic models of the AGN/ICM
interaction
\citep{1997MNRAS.284..981C,2001ApJ...554..261C,2001MNRAS.325..676B,2002Natur.418..301B,2002MNRAS.332..271R,2003MNRAS.339..353B,2004MNRAS.348.1105O,2004MNRAS.350L..13O,2004MNRAS.355..995D,2004ApJ...601..621R,2005A&A...429..399Z}.
Some groups have also begun initial explorations of the effects of
magnetic fields \citep{2004ApJ...601..621R,2005ApJ...624..586J},
plasma processes \citep{2004ApJ...611..158R,2005MNRAS.357..242R},
feedback prescriptions~\citep{2006ApJ...645...83V}, and realistic
(cosmological) background motions~\citep{2006MNRAS.373L..65H}.

Collectively, this body of simulation work has allowed us to explain
many of the observed features of AGN/ICM interactions, investigate how
AGN induced flows mix metals within the ICM core, and study entropy
injection and heating of the ICM.  However, most of the current sets
of simulation are performed for a limited set of jet parameters and
hence it is unclear how to generalize the results to the population of
as a whole.  In this paper, we present a moderately large set of
(axisymmetric) high-resolution hydrodynamic simulations of jet/ICM
interactions that survey a wide range of jet powers and jet
velocities.  We study the morphology of the radio-galaxy as well as
the injection of energy and entropy into the ICM as a function of the
jet properties.  In Section~\ref{sec:setup}, we will discuss the
details of our setup and the code used.  In
Section~\ref{sec:analysis}, we will present the analysis of our
simulations and their classifications.  In Section~\ref{sec:disc} we
will discuss the results, followed by conclusions in
Section~\ref{sec:conc}.

\section{Simulation Setup}
\label{sec:setup}

Our goal is to model a relaxed cluster and its interaction with a
central radio galaxy that produces back-to-back jets.  To do this, we
start out with an initial cluster that is both spherically symmetric
and isothermal, with the (adiabatic) sound speed $c_s=1$ everywhere.
The initial cluster gas is assumed to follow a $\beta$-model density
profile,
\begin{equation}
  \label{eq:beta}
  \rho(r)=\frac{1}{[1+(\frac{r}{r_0})^2]^{3/4}},
\end{equation}
where the core radius is $r_0=2.0$ in code units.

A static gravitational potential,
\begin{equation}
  \label{eq:grav}
  \Phi=\frac{c^2_s}{\gamma}\ln(\rho),
\end{equation}
is set to make the gas initially in hydrostatic equilibrium and is
assumed to remain fixed in time.  This is equivalent to a potential
set entirely by stationary dark matter that dominates the system, so
the self gravity of the gas is ignored.  The intracluster gas, while
containing a large fraction of the baryons in the cluster, is indeed
not a significant contributor to the overall cluster mass.

We use spherical polar coordinates ($r$, $\theta$, $\phi$) and impose
symmetry in the $\phi$ direction ($\partial/\partial\phi=0$).  The
radial coordinate varied between $0.1$ and $30.0$ in scaled code
units.  The comparison between code units and physical units is given
in Section~\ref{sec:real}.  The values of $\theta$ were allowed to
vary between $0$ and $\pi$ (i.e., we are modeling the full sphere).
Since we have the full range of angles, this allows us to have
back-to-back jets.  Thus we can model interaction between the
backflows from the jets rather than the commonly employed technique of
imposing reflecting boundary conditions on the $\theta=\pi/2$ plane
(e.g., see ~\citet{2007ApJ...656L...5S,2006ApJ...645...83V}).  We
included second order differences in grid spacing in the $\theta$
direction for each hemisphere,
giving us two realizations of the jet/cocoon structure per
simulations.  In no cases were there major differences between the two
sides, confirming that the details of the griding do not affect our
results.  A small circle around $r=0$ was excluded from the
computational domain to avoid the coordinate singularity.  

All simulations were run with a $n_r\times n_\theta=1200\times600$
grid.  In the radial direction, a ratioed grid was used, with
successive cells increasing by a factor of $1.003$.  The $\theta$
direction also used a ratioed grid increasing by a factor of $0.998$
from $0$ to $0.5\pi$ and then decreasing by a factor of $1.002$ from
$0.5\pi$ to $\pi$.  This provides the greatest number of zones near
the center and near the two jet axes while providing the second-order
differences in the $\theta$ direction mentioned above.

Clearly jets are three dimensional structures.  However, there is a
long history of two-dimensional modeling of jets which supports the
usefulness of two dimensional jets~\citep{2002ApJS..141..371C,
2001ApJ...554..261C, 2001JKAS...34..329M, 2002MNRAS.332..271R}.  There
are however several aspects of the dynamics that cannot be captured in
two dimensions.  There is no way to have realistic random motions in
the background.  This lack of random motions produces lobes which are
far more regular and symmetric than those found in any real
source. The two dimensional assumption also enhances mixing between
the high entropy material and ambient material. This is because
non-axisymmetric Kelvin-Helmholtz (KH) and Rayleigh-Taylor (RT) modes 
are clearly not possible in an axisymmetric system. Without
non-axisymmetric modes, all of the mixing must be done by the
axisymmetric modes, which are more efficient at mixing
\citep{2002MNRAS.332..271R}. Fortunately, this will be most
problematic only at late times when there has been enough time
for significant mixing to occur.

The hydrodynamic evolution of the jet/ICM system was followed using
the ZEUS-3D code~\citep{1992ApJS...80..753S, 1992ApJS...80..791S}.
ZEUS is a fixed grid Eulerian code, explicit in time and second order
accurate (when using van Leer advection as we do).  Artificial
viscosity is used to handle shocks.  All simulations were run as ideal
hydrodynamic cases, neglecting magnetic fields and self gravity.  ZEUS
solves the standard equations of hydrodynamics,
\begin{equation}
  \label{eq:hd1}
  \frac{D \rho}{D t}+\rho\nabla\cdot v = 0,
\end{equation}

\begin{equation}
  \label{eq:hd2}
  \rho\frac{D v}{D t}=-\nabla P -\rho\nabla\Phi,
\end{equation}

\begin{equation}
  \label{eq:hd3}
  \rho\frac{D }{D t} \left( \frac{e}{\rho} \right) =-P\nabla\cdot v - \Lambda,
\end{equation}
where
\begin{equation}
  \label{eq:hd4}
  \frac{D }{D t}\equiv\frac{\partial}{\partial t} + v\cdot\nabla.
\end{equation}
Radiative cooling, represented by the $\Lambda$-term in
Equation~\ref{eq:hd3}, is
neglected in order to concentrate on the hydrodynamic evolution and
the thermodynamic response of the ICM to the jets.

\subsection{Modeling the jets}
\label{sec:jet}
Two, back-to-back jets were injected into the center of the model
cluster.  This was achieved with an inflow boundary condition on the
inner radial boundary.  All the jets were highly supersonic with
respect to the ambient sound speed and moderately to highly supersonic
with respect to their internal sound speeds.  Direct injection allows
us to keep the details of jet acceleration and central engines outside
of the grid, so we are able to solve the equations of hydrodynamics
consistently {\it within our computational grid}.  

The injected jets are initially conical with half-opening angles of
$15^\circ$, and are injected in pressure balance with the ICM core.
The drop of internal pressure associated with subsequent jet expansion
leads to external, pressure driven recollimation and internal shocks,
ultimately resulting in some new, self-consistently determined opening
angle.  Given these (fixed) assumptions, the properties of a given jet
are parameterized by the kinetic luminosity $L_{\rm kin}$ and internal
Mach number of the injected jet matter.  These are related to the
injection density $\rho$, pressure $p$ and velocity $v$ by
\begin{eqnarray}
  \mathcal{M}_{jet}=\frac{v_j}{c_s}&=v_j\sqrt{\frac{\rho}{\gamma(\gamma-1)e}},\\
  L_{kin}&=\frac{1}{2}\rho_j v_j^3 \pi r_{in}^2.
  \label{eq:lkin}
\end{eqnarray}
where $r_{\rm in}$ is the radius of the ``injection nozzle'' for the
jet.  The jet parameters for the base run ($L_{\rm kin}=26.47$ and
$\mathcal{M}_{\rm jet}=10.55$; $\rho_j=0.01$ and $v_j=105.5$) are the same
as the run presented in \citet{2002MNRAS.332..271R}.

At total of 26 models were run, where only the kinetic luminosity
and/or the Mach number of the jets were varied.  In two of these
simulations, the jet lifetimes were also varied.  The full list of
simulations is given in Table~\ref{t:sims}.  Internal Mach numbers in
the range 3.7--21.1, and kinetic luminosity in the range 6.6---68.8
were explored.  The choice of parameters used was based on raising or
lowering the Mach number and/or kinetic luminosity by a factor of two
from one of the previous runs.


\begin{table}

\caption{Parameters defining the 26 simulations presented in this
paper.  Also given are the (visual) morphological classification of
the resulting structure, as judged at time $t=2.0$.  See text for more
details.}
\begin{center}
\begin{tabular}{c c c c c } \hline \hline
Run & Jet Active Time & $L_{kin}$ & $\mathcal{M}_{jet}$ & Morphology \\ \hline
1 & 1.0 & 26.47 & 10.55 & Cocoon\\
2 & 1.0 & 52.94 & 21.1 & Cocoon\\
3 & 1.0 & 26.43 & 15.89 & Cocoon\\
4 & 1.0 & 26.31 & 7.39 & Cocoon\\
5 & 1.0 & 68.97 & 10.46 & Cocoon\\
6 & 1.0 & 23.32 & 10.54 & Cocoon\\
7 & 1.0 & 52.94 & 10.55 & Cocoon\\
8 & 0.5 & 26.47 & 10.55 & Cocoon\\
9 & 1.5 & 26.47 & 10.55 & Cocoon\\
10 & 1.0 & 8.91 & 6.47 & Cocoon\\
11 & 1.0 & 17.91 & 9.48 & Cocoon\\
12 & 1.0 & 19.98 & 15.0 & Cocoon\\
13 & 1.0 & 31.09 & 20.02 & Cocoon\\
14 & 1.0 & 13.24 & 21.1 & Non-Cocoon\\
15 & 1.0 & 13.23 & 10.55 & Non-Cocoon\\
16 & 1.0 & 26.47 & 21.1 & Non-Cocoon\\
17 & 1.0 & 6.62 & 10.55 & Non-Cocoon\\
18 & 1.0 & 9.96 & 8.99 & Non-Cocoon\\
19 & 1.0 & 21.01 & 13.0 & Non-Cocoon\\
20 & 1.0 & 30.1 & 18.02 & Non-Cocoon\\
21 & 1.0 & 24.98 & 19.0 & Non-Cocoon\\
22 & 1.0 & 52.99 & 5.28 & Unresolved\\
23 & 1.0 & 26.44 & 3.71 & Unresolved\\
24 & 1.0 & 26.43 & 5.27 & Unresolved\\
25 & 1.0 & 10.05 & 5.01 & Unresolved\\
26 & 1.0 & 12.15 & 6.54 & Unresolved\\
\hline
  \end{tabular}
\end{center}
\label{t:sims}
\end{table}

\subsection{Comparison to Real Units}
\label{sec:real}
Throughout this paper, quantities are quoted in dimensionless code
units unless explicitly stated otherwise.  Since we are solving the
equations of ideal hydrodynamics (Equations~\ref{eq:hd1}-\ref{eq:hd4})
with no additional physics added, there is no unique set of physical
scales to our problem.  Each simulation may therefore be compared to
several different sets of physical scales.  Following Reynolds et
al. (2002), we shall quote two scalings for our simulations (although
there is actually an entire three-parameter family of scalings).  For
our rich cluster scaling, we set the core radius $r_0=100\,{\rm kpc}$,
meaning one code unit in r is equal to $50\,{\rm kpc}$.  Such clusters
are hot, with sound speed, $c_s=1000\,{\rm km}\,{\rm s}^{-1}$. This,
along with the typical central number density $n_0=0.01\,{\rm
cm}^{-1}$, gives a time unit of $50$\,Myrs and a kinetic luminosity
unit of $L_{kin}=3.5\times10^{44}{\rm erg}\,{\rm s}^{-1}$.  We can
also consider a poor cluster scaling, $r_0=50\,{\rm kpc}$,
$c_s=500\,{\rm km}\,{\rm s}^{-1}$, and $n_0=0.1\,{\rm cm}^{-1}$.
These give a length unit of $25$\,kpc, a time unit of $10$ Myrs and a
kinetic luminosity unit of $L_{kin}=4.4\times10^{42}\,{\rm erg}\,{\rm
s}^{-1}$.

\section{Analysis and Results}
\label{sec:analysis}

The primary diagnostic for distinguishing injected jet material from
ambient and shocked-ambient material is the specific entropy index,
\begin{equation}
  \label{eq:ent}
  \sigma=\frac{P}{\rho^{\gamma}}.
\end{equation}
The ambient material (both undisturbed and shock heated) has a low
specific entropy index (which ranges from $0.6$ to $9$) and the
injected jet material has a specific entropy index on the order of
$1000$ for our base simulation (run 1).  The specific entropy of
injected material does vary with jet power, going from near $1$ for
the weakest jets to $10^7$ for the most powerful.  With only two ways
to change entropy (increase with shocks and increase or decrease
through numerical mixing), specific entropy generally provides a
powerful way to distinguish injected (and shocked) material from the
background gas.  A cutoff of $\sigma=10$ was set to separate jet
material from ambient material.  This is close to the highest ambient
value so we are not likely to artificially miss any material that
should count as part of the cocoon/high entropy material.  We use the
same entropy-index cutoff for all simulations to allow for direct
comparisons to be made.  This value was chosen to avoid counting
ambient material which has the same range of values in all
simulations.  In the case of the weakest jets, this entropy index
cutoff cannot pick out the pre-shocked jet material.  Once the jet
material passes through the shock terminal shock, however, it is
always clearly distinguished from ambient material by a $\sigma=10$
cutoff.  For a further discussion of the entropy cutoff diagnostic,
see~\citet{2002MNRAS.332..271R}.

\subsection{Morphological Classification}
\label{sec:morph}

Based on the entropy maps, the simulations were visually placed into
three classes.  Simulations with a distinct, well-defined monolithic
cocoon of jet-originating plasma were labeled as ``Cocoon bounded
sources'' (or simply ``cocoon sources'').  We consider a cocoon to be
a jet inflated region of high entropy material with a coherent
structure that is well separated from the low entropy background and
relatively safe from mixing (until late times).  Simulations which
showed faint, wispy plumes of jet-originating material instead of well
defined cocoons were labeled ``Non-Cocoon bounded sources''
(``non-cocoon sources'').  The final group were those simulations that
do not appear sufficiently resolved for an accurate classification.
In the discussion below, we shall abbreviate the term
``jet-originating material'' to ``jet material''.

\begin{figure*}
  \centering
  \epsscale{0.4}
  \plotone{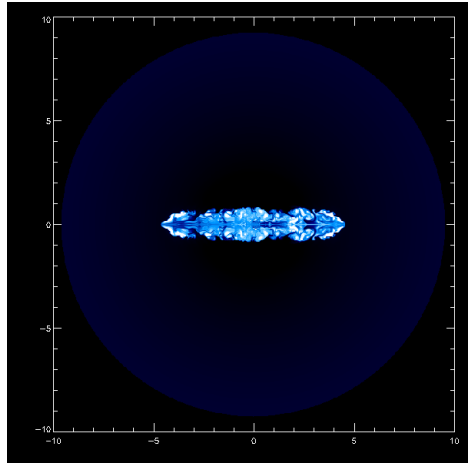}
  \caption{Entropy map of central regions of typical cocoon bounded
    source (run 1) at t=2.0.}
  \label{fig:coc}
\end{figure*}

\begin{figure*}
  \centering
  \epsscale{0.4}
  \plotone{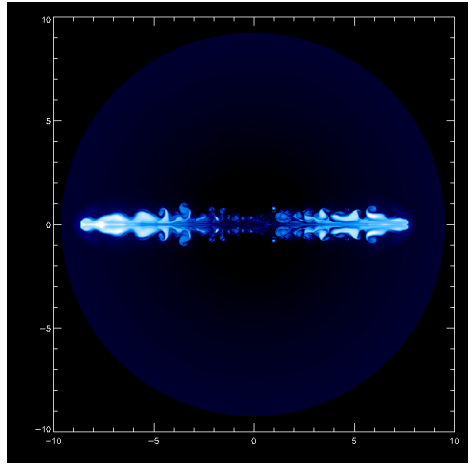}
  \caption{Entropy map of central regions of typical non-cocoon
    bounded source (run 21) t=2.0.} 
  \label{fig:noncoc}
\end{figure*}

\begin{figure*}
  \centering
  \epsscale{0.4}
  \plotone{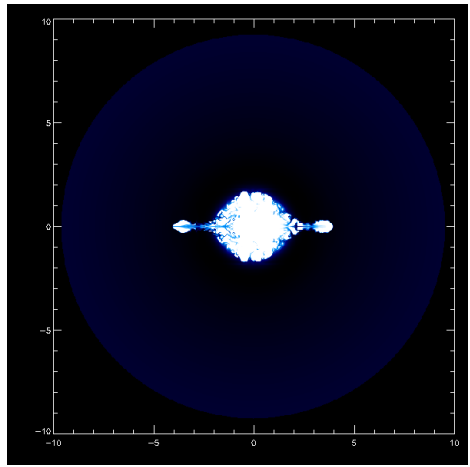}
  \caption{Entropy map of central regions of typical unresolved source
    (run 23) at t=2.0.}
  \label{fig:unres}
\end{figure*}

\begin{figure*}
  \centering
  \epsscale{0.6}
  \plotone{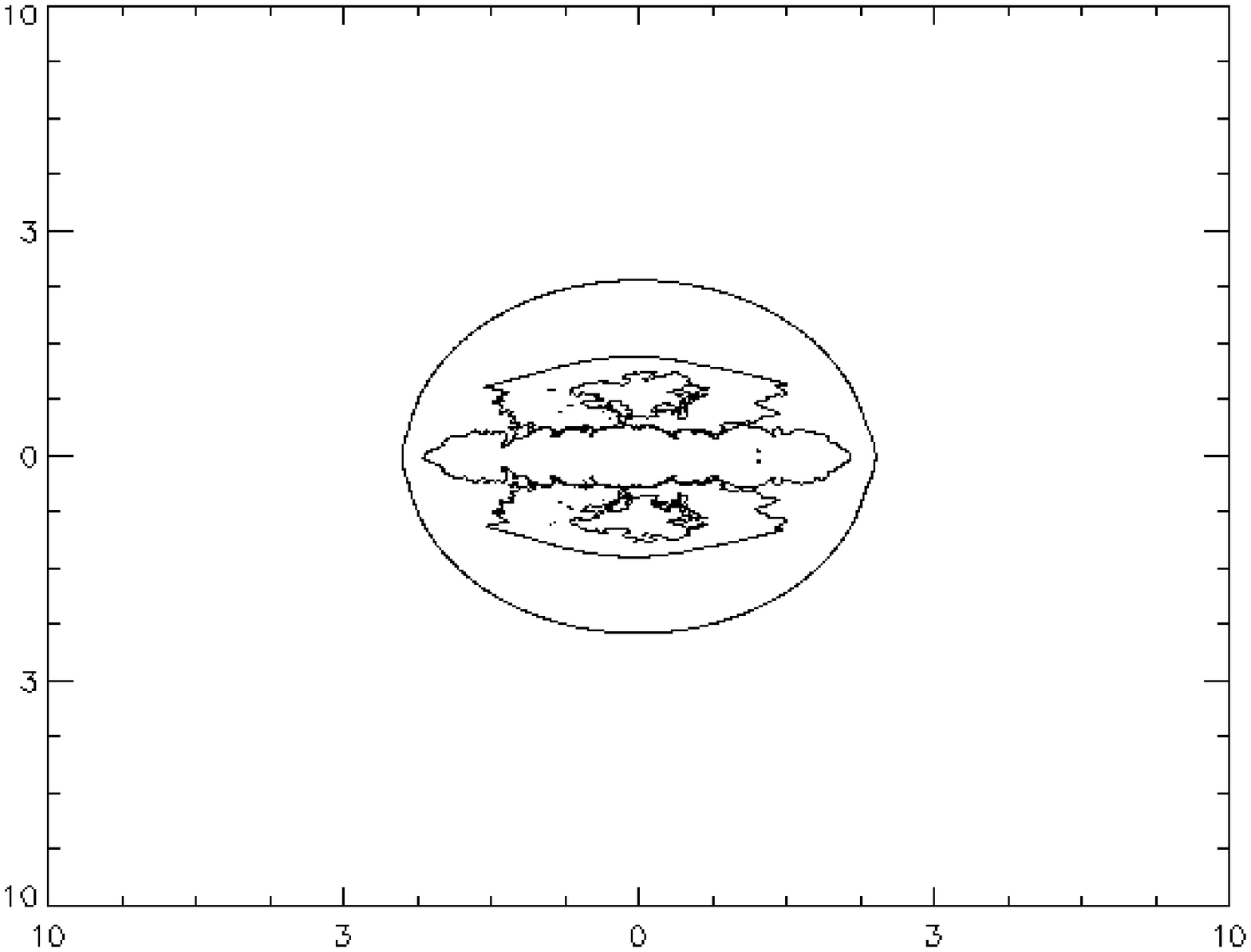}
  \caption{Density contours for inner half of cocoon bounded source
  (run 1) at t=2.0}
  \label{fig:coc-cont}
\end{figure*}

\begin{figure*}
  \centering
  \epsscale{0.6}
  \plotone{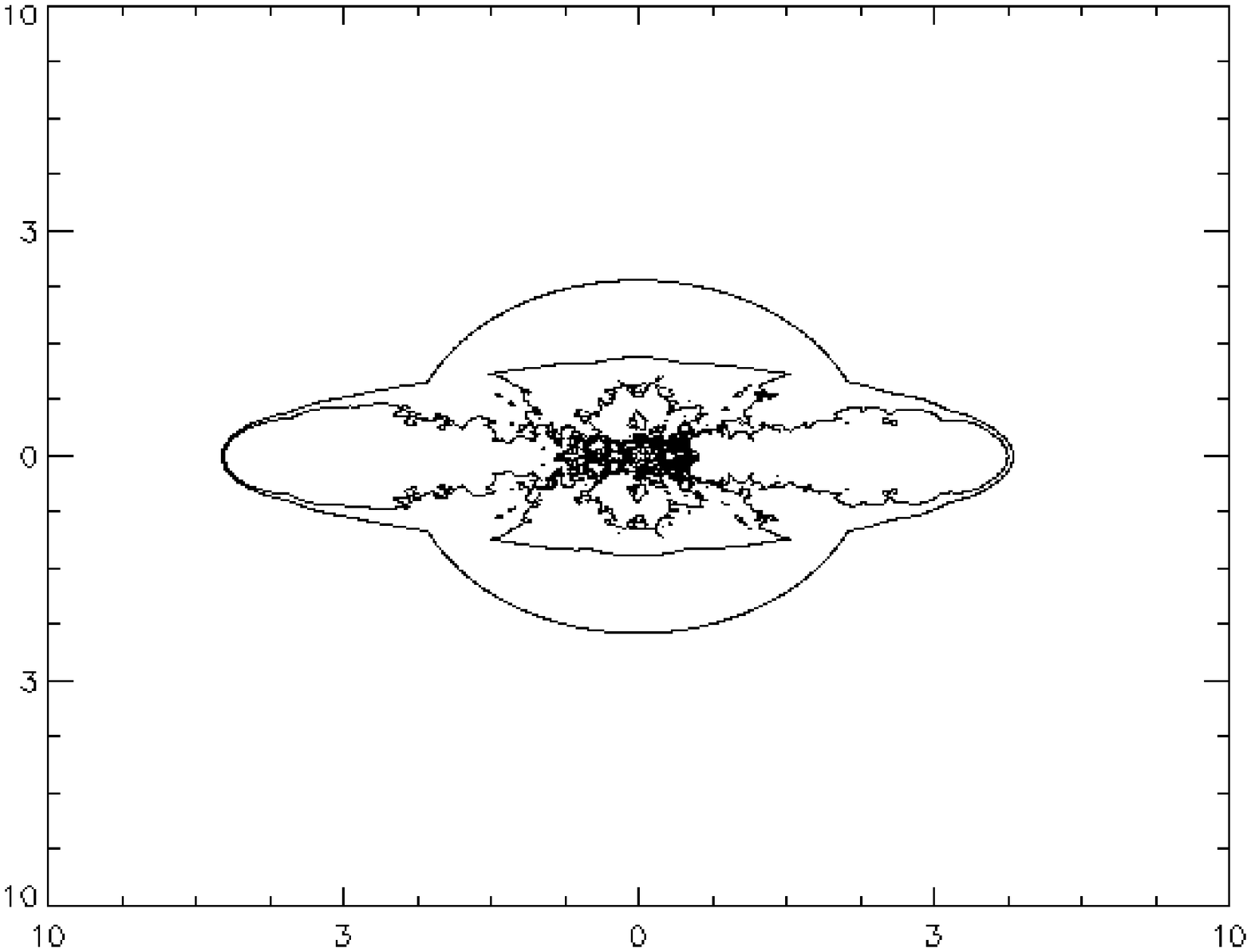}
  \caption{Density contours for inner half of non-cocoon bounded
  source (run 21) at t=2.0}
  \label{fig:ncoc-cont}
\end{figure*}

An example of a typical cocoon source can be seen in
Figure~\ref{fig:coc}.  Figure~\ref{fig:noncoc} shows a typical
non-cocoon source.  A poorly resolved simulation can be seen in
Figure~\ref{fig:unres}.  We also observe that it is possible to
separate the cocoon-bounded from the non-cocoon bounded sources by use
of density contours.  Figures~\ref{fig:coc-cont}
and~\ref{fig:ncoc-cont} show density contours for a cocoon bounded and
non-cocoon bounded source respectively.  Simulations in the cocoon
category appear to have the high entropy jet material in a distinct
region separated from the background by a solid density contour and
remain that way until fairly late times.  Non-cocoon sources do not
have such distinctions and the jet material seems mixed in with the
background even early on.

These classifications were performed at the same time for all
simulations, $t=2.0$.  At this time, the structures have evolved
passively for the same amount of time that they were driven by the
jet.  However, our classification does not appear to depend
sensitively on the choice of time, although there are some
limitations.  It the very earliest stages of the active jet phase
($t<0.2-0.4$), the jets are still pushing out from the inner boundary
of the simulation (i.e., the morphology is strongly determined by the
artificial aspects of the simulation) and classification is not
meaningful.  At late times ($t>10-15$), mixing has disrupted all of our
simulated sources and, again, eliminated any morphological
distinction.

\begin{figure*}
  \centering
  \epsscale{0.6}
  \plotone{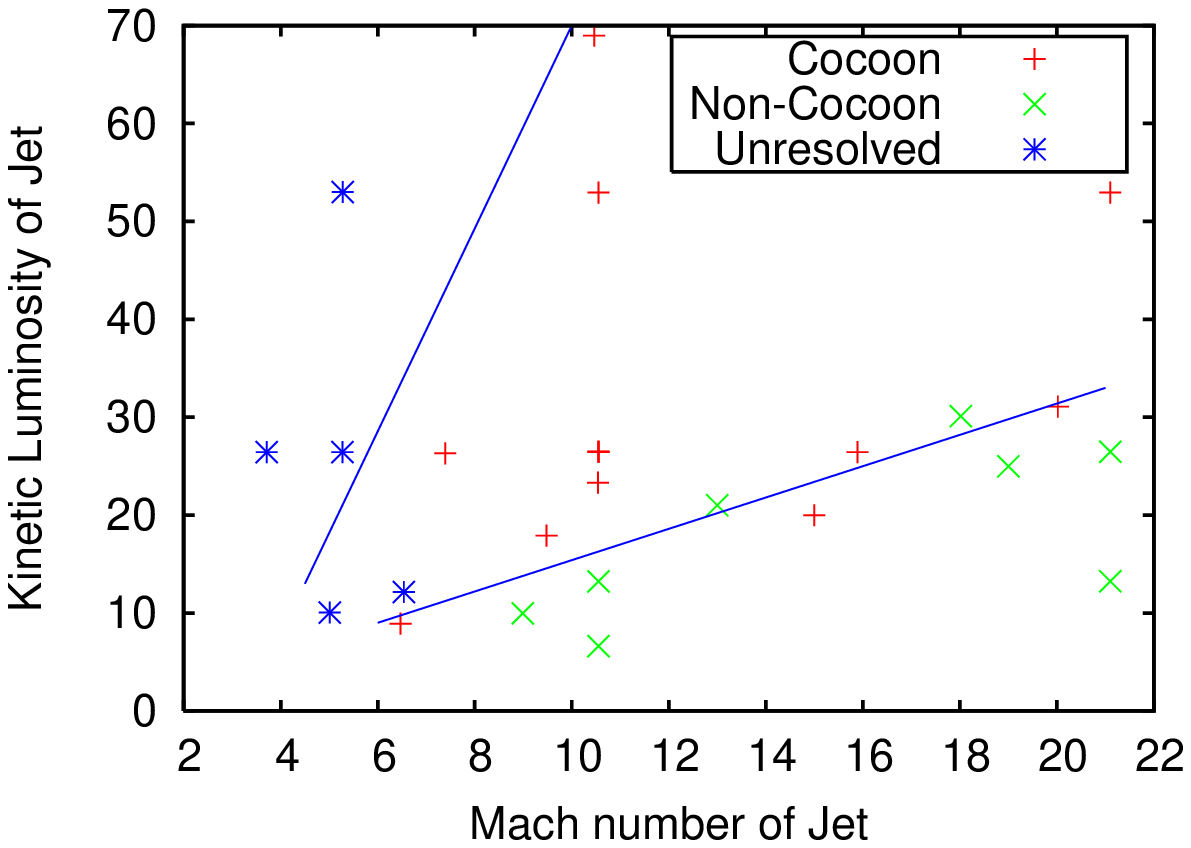}
  \caption{Morphology vs. jet parameters.  Lines separating the
  different regions were done by eye.  Although the
  lines between the different regions are not perfect, there appear to
  be regions in which different classes are clearly excluded.}
  \label{fig:pspace}
\end{figure*}

\begin{figure*}
  \centering
  \epsscale{0.6}
  \plotone{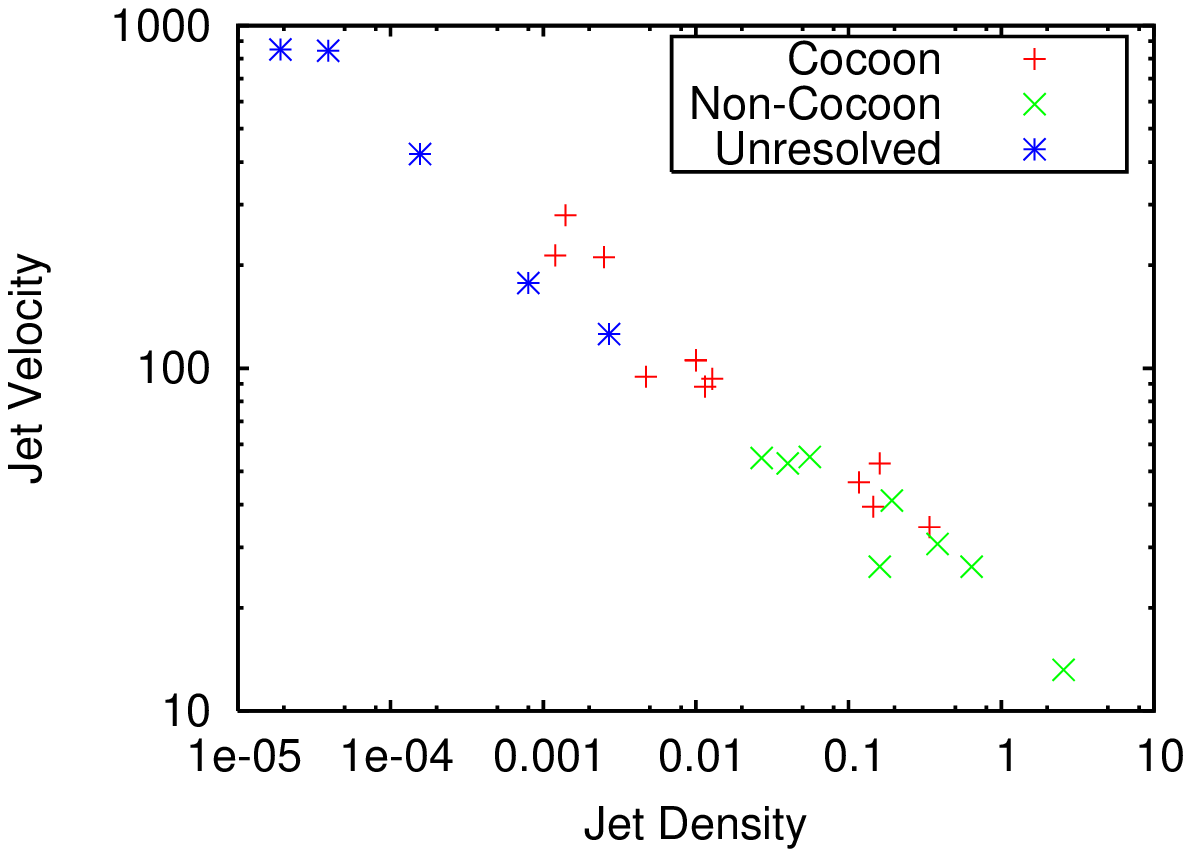}
  \caption{Morphology vs. jet parameters.  On the density-velocity
    plane, we see the previously distinct regions as more of a
    continuum moving from one type of source to the next.}
  \label{fig:pspace2}
\end{figure*}

Using this classification, we can study how morphology depends on
$L_{\rm jet}$ and $\mathcal{M}_{\rm jet}$.  Our results are shown in
Fig.~\ref{fig:pspace}.  We see three separate regions on this
parameter space.  At the high kinetic luminosities and lower Mach
numbers, the runs are unresolved.  From this point on, the unresolved
sources were excluded from all analysis.  Low luminosity sources seem
to fall in the non-cocoon category, almost regardless of Mach number
(except that at the high Mach number they can have a higher kinetic
luminosity).  The middle region contains the cocoon bounded sources.
In Fig.~\ref{fig:pspace2}, we show the parameter space again, on a
density-velocity plot.  Although the overlap between the morphologies
is more apparent in this parameter space, we can see that there
appears to be a continuum of morphologies with the fastest and
lightest jet producing unresolved sources at one end and the slow
heavy jets producing non-cocoon sources.  Again, the cocoon sources
occupy the middle ground between the other two.  A discussion of the
physical origin of the cocoon will be reserved for the Discussion
(Section~\ref{sec:disc}).

As revealed by Fig.~\ref{fig:pspace}, the line separating cocoon
bounded and non-cocoon bounded sources on the $(L_j,{\cal M}_{\rm
jet})$ is approximately linear (i.e., $L_j\propto {\cal M}_{\rm
jet}$).  This can be understood from the analytic models of cocoon
expansion from \citet{1989ApJ...345L..21B}.  Equation (4) of
\citet{1989ApJ...345L..21B} states the the pressure of a cocoon is
given by
\begin{equation}
p_c\sim \frac{(L_{\rm kin}v_j\rho_aA_h)^{1/2}}{A_c}
\end{equation}
where $A_h$ is the area of the working surface over which the jet
deposits its momentum, $A_c$ is the cross-sectional area of the
cocoon, and $\rho_a$ is the density of the ambient ISM/ICM.  Well
defined cocoons can only be sustained if $p_c$ exceeds the pressure of
the ambient medium, $p_a$.  Noting that the axial ratio of the cocoons
are approximately constant (i.e., $A_h/A_c\sim {\rm constant}$), and
that the density and pressure of our model ambient ICM atmosphere
outside of the ICM core in the same in all simulations and drops off
as $r^{-3/2}$, the condition that $p_c>p_a$ can be written as the
condition 
\begin{equation}
(L_{\rm kin}v_j)^{1/2}>\Upsilon_{\rm crit},
\label{eqn:cocoon_pressure}
\end{equation}
where $\Upsilon_{\rm crit}$ is a weak function of $r$ ($\Upsilon
\propto r^{1/4}$) and hence time.  Since the jets are injected with
fixed initial pressure, it can be shown that initial jet velocity
depends on the kinetic luminosity and initial internal Mach number
according to $ v_j\propto L_{\rm kin}/{\cal M}_{\rm jet}^2$.  Thus, we
can use eqn.~\ref{eqn:cocoon_pressure} to see that the line
separating cocoon from non-cocoon sources (i.e., where the cocoon just
comes into pressure balance with the ambient ICM) can be written as
$L_{\rm jet}\propto \Upsilon {\cal M}_{\rm jet}$.  


\subsection{Mass Distribution and Energetics}
\label{sec:energetics}

To examine the mass distribution and energetics during the
simulations, several quantities were calculated at each output time.
The total mass of jet material at each time is given by:
\begin{equation}
  M_{cocoon}=\int_C \rho d V
  \label{eq:mcoc}
\end{equation}
where $C$ is the region with $\sigma>10$.  The internal, kinetic, and
potential energies were calculated for both the jet and the ambient
material,
\begin{eqnarray}
  E_{int}^{C,A}&=\frac{1}{\gamma-1}\int_{C,A} P d V,\\
  E_{kin}^{C,A}&=\frac{1}{2}\int_{C,A} \rho v^2 d V,\\
  E_{pot}^{C,A}&=-\int_{C,A}\rho\Phi d V.
\end{eqnarray}
where $A$ is the region with $\sigma\le 10$.  In all cases, the
integral is taken over only the region of interest (i.e., cocoon or
ambient only, as determined by the entropy cutoff).  We then subtract
the initial energies in order to derive the change of energy $\Delta
E_{int}^{C,A}$, $\Delta E_{kin}^{C,A}$, and $\Delta E_{pot}^{C,A}$.

\begin{figure*}
  \centering
  \epsscale{0.35}
  \plotone{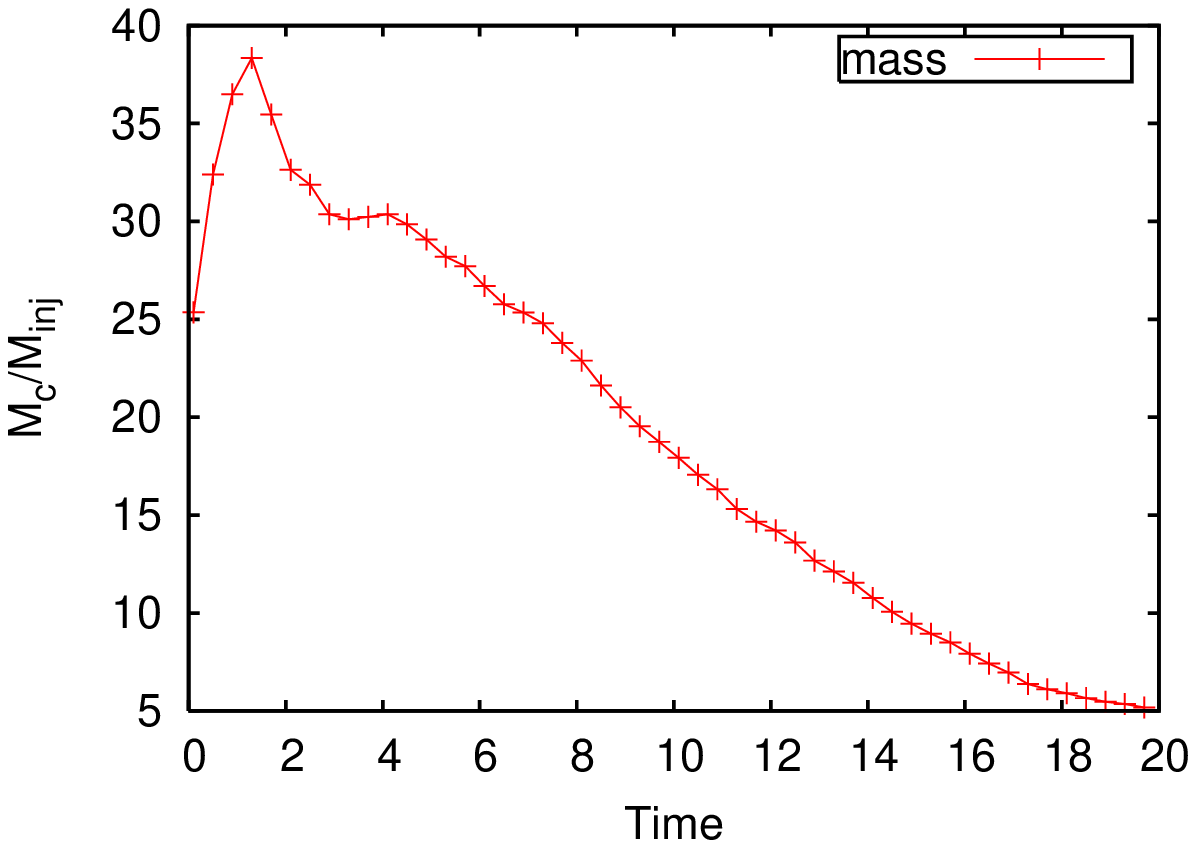}
  \plotone{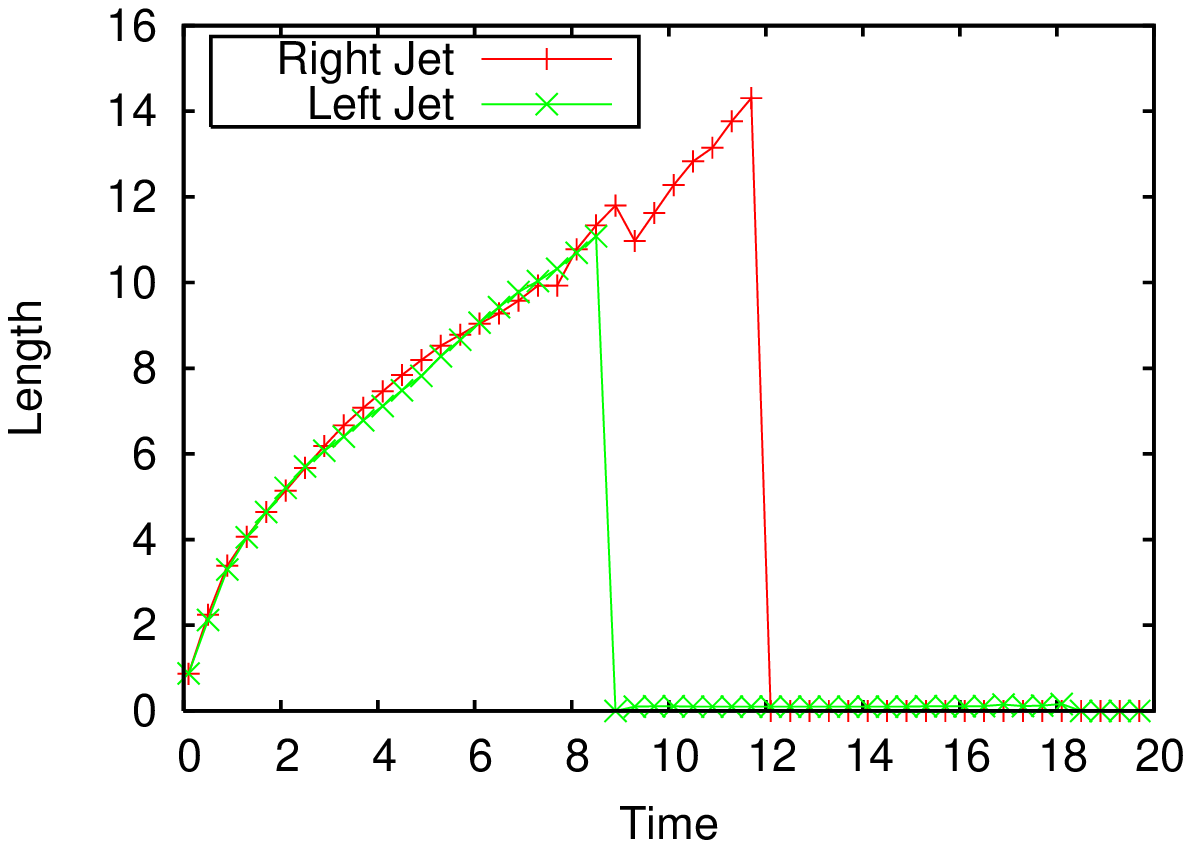}

  \plotone{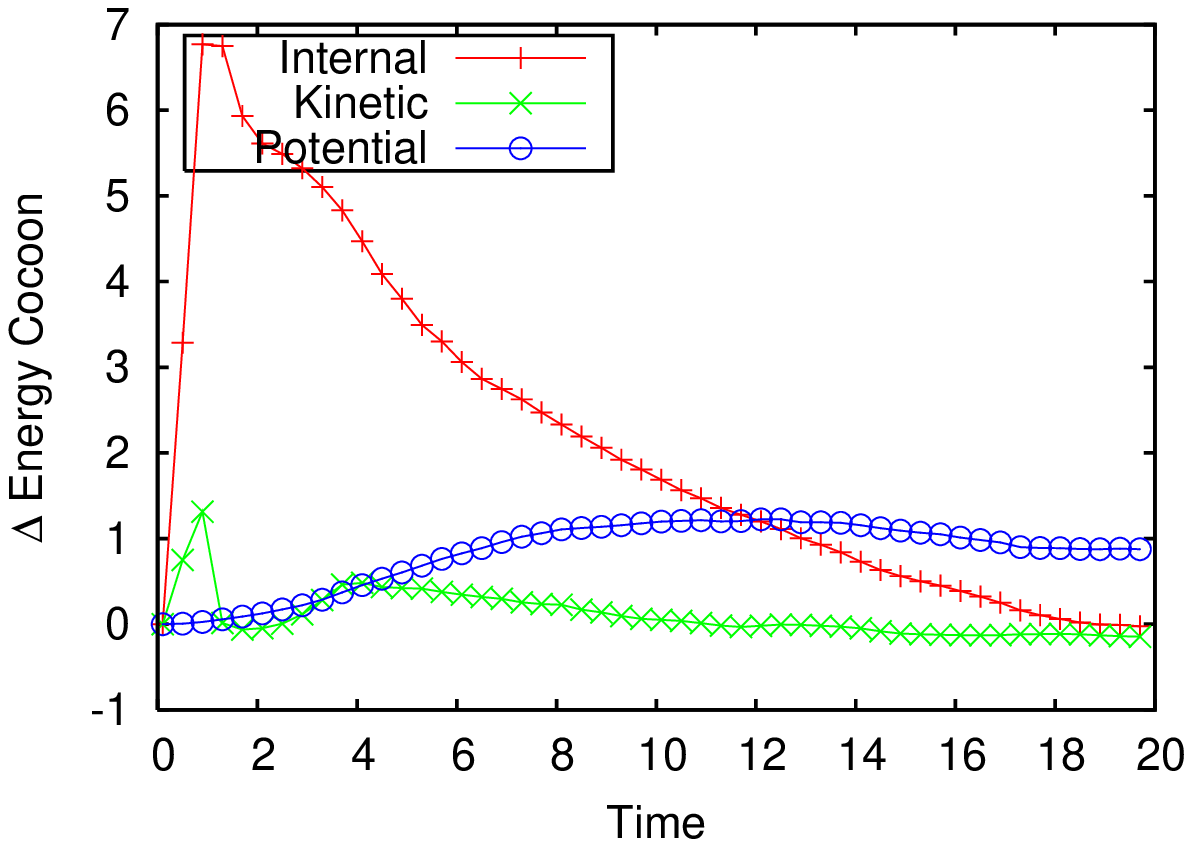}
  \plotone{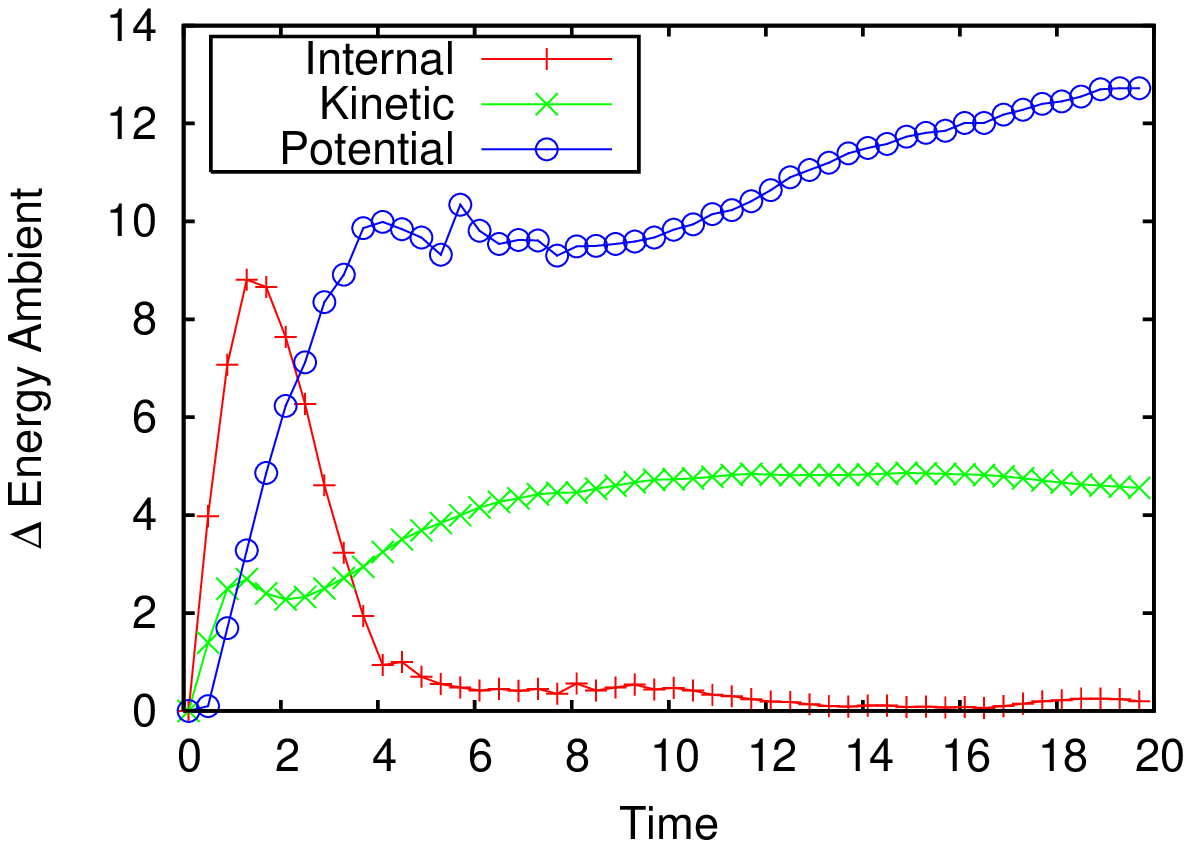}
  \caption{Mass and Energy vs. time for cocoon bounded simulation
    (run 1).
    Starting from the top left and moving clockwise, the panels are:
    Mass of the cocoon material divided by the injected
    mass, the maximum radial extent of the high entropy material, the
    change in internal, kinetic, and potential energies for the
    ambient, low entropy 
    material compared to the initial state, and the same energy
    changes for the cocoon material.}  
  \label{fig:cmasses}
\end{figure*}

\begin{figure*}
  \centering
  \epsscale{0.35}
  \plotone{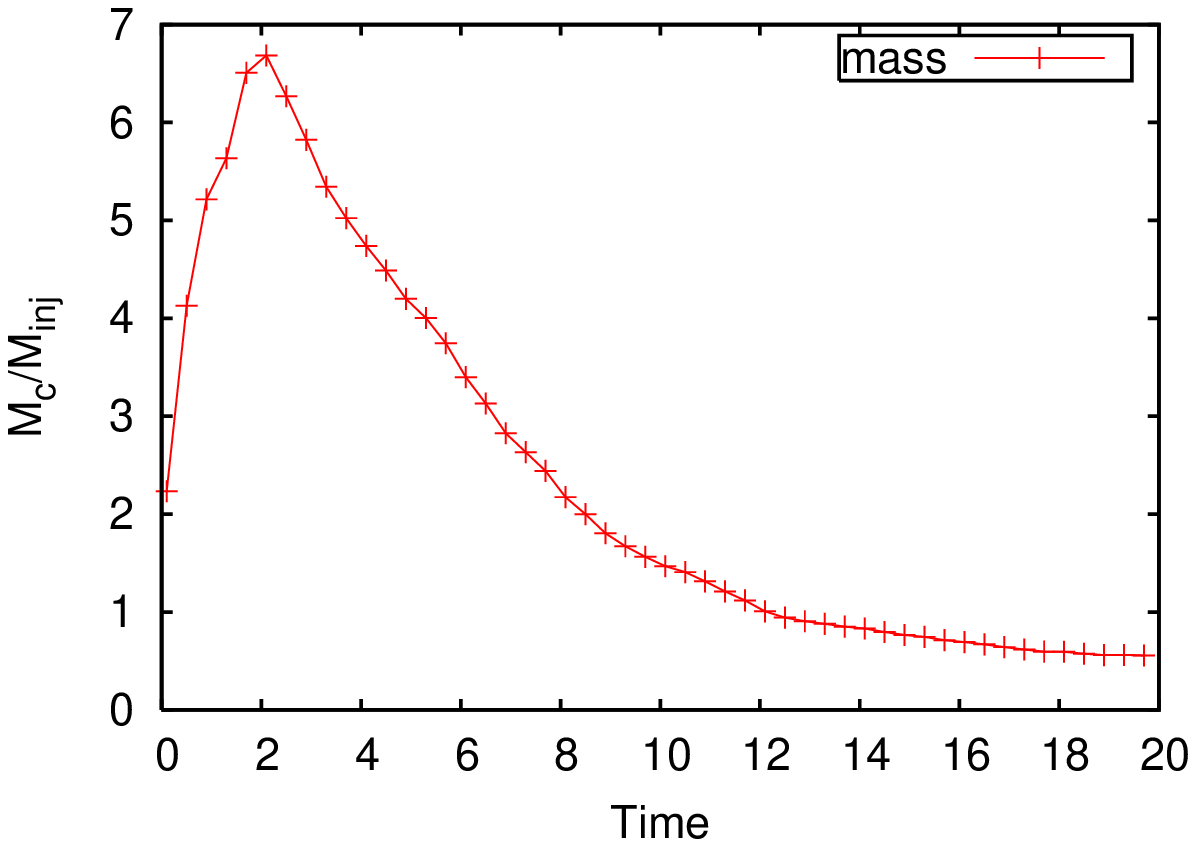}
  \plotone{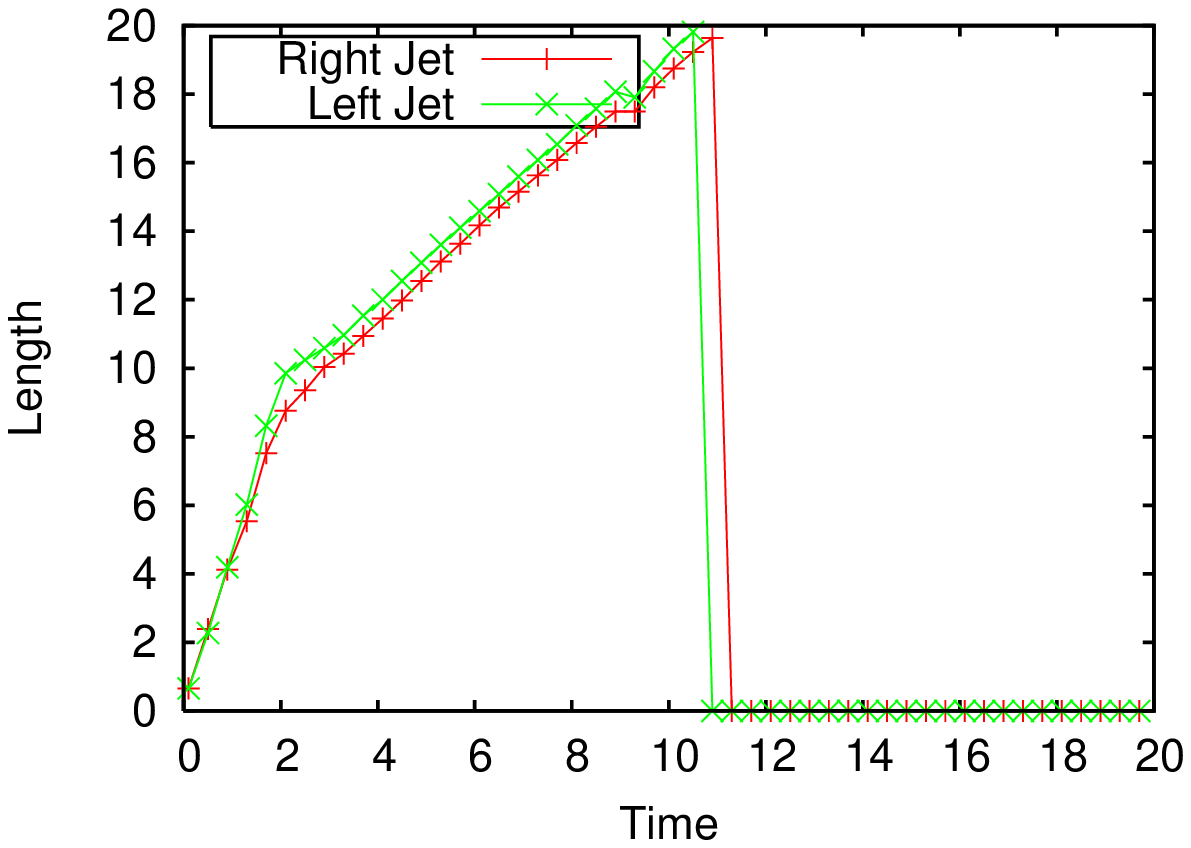}

  \plotone{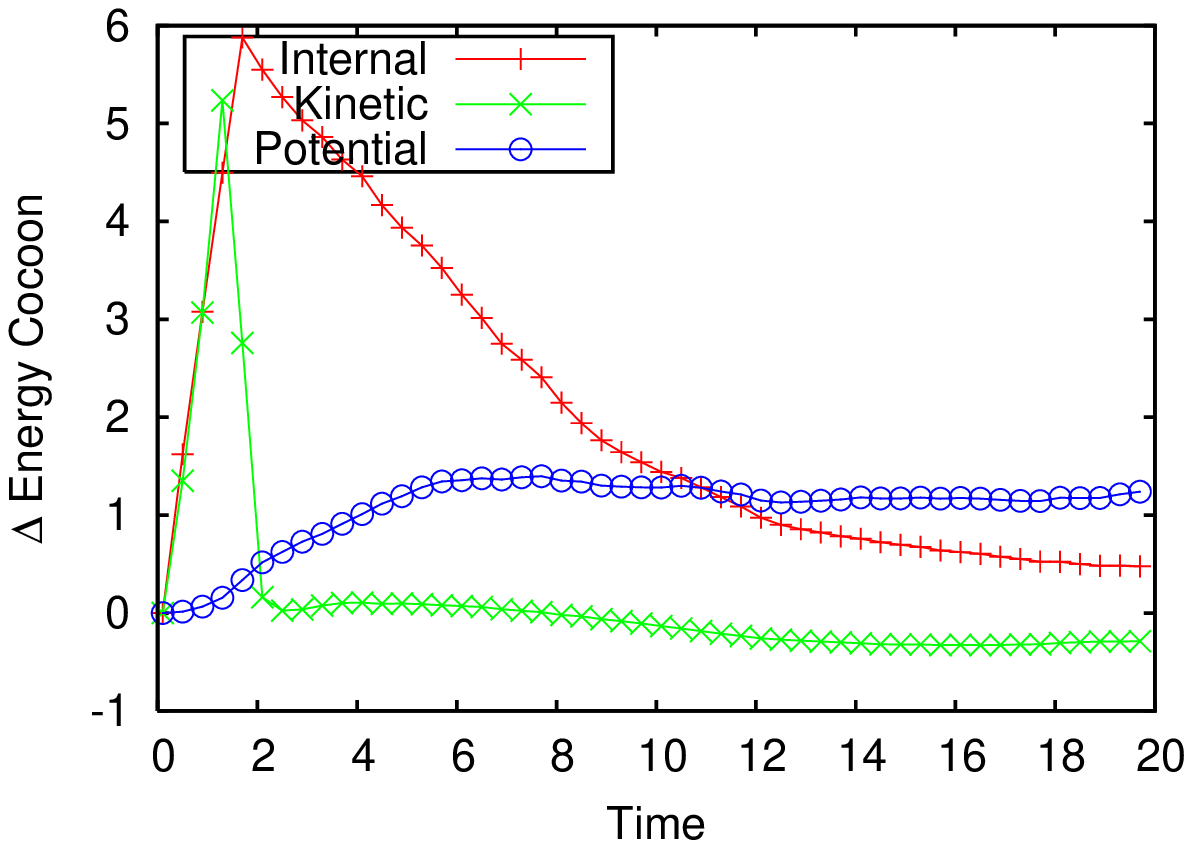}
  \plotone{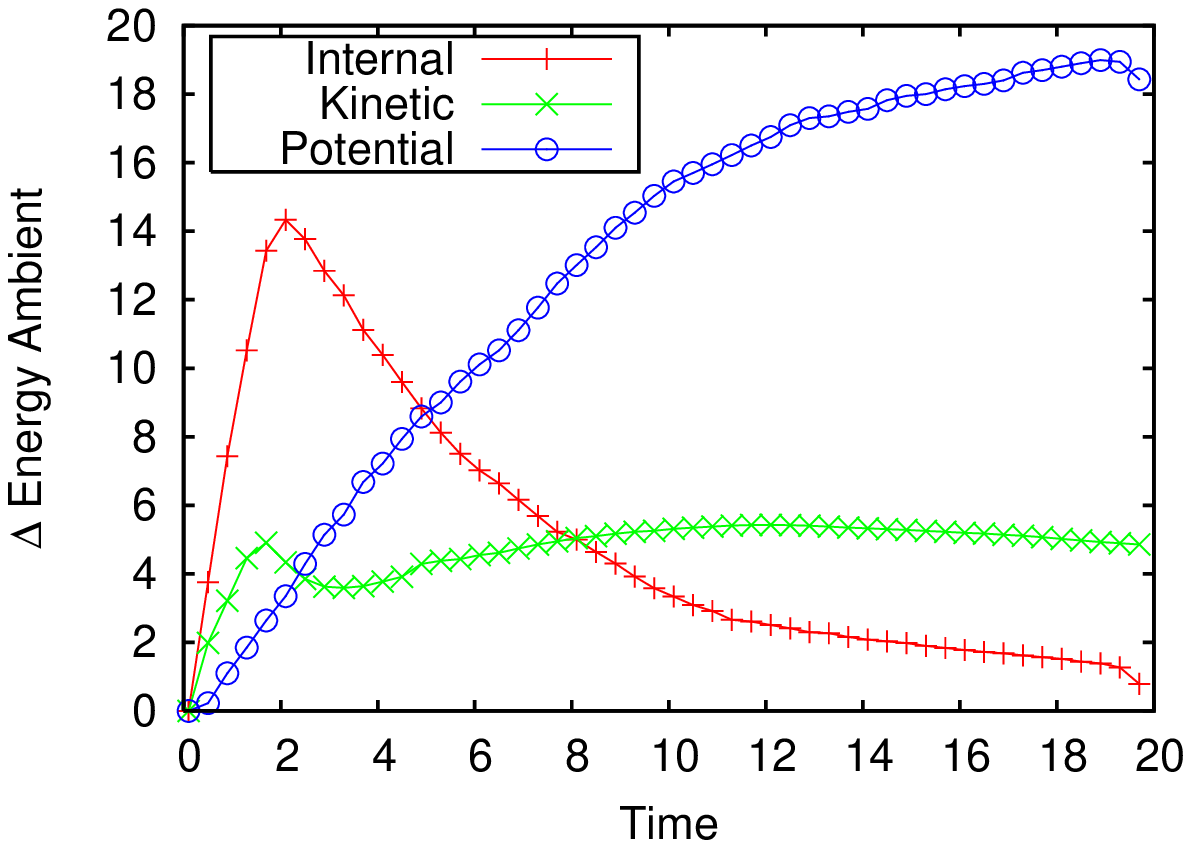}
  \caption{Mass and Energy vs. time for non-cocoon bounded
    simulation (run 21).
    Starting from the top left and moving clockwise, the panels are:
    Mass of the cocoon material divided by the injected
    mass, the maximum radial extent of the high entropy material, the
    change in internal, kinetic, and potential energies for the
    ambient, low entropy 
    material compared to the initial state, and the same energy
    changes for the cocoon material.}
  \label{fig:ncmasses}
\end{figure*}

The results of these calculations for a representative cocoon bounded
simulation are shown in Figure~\ref{fig:cmasses}.
Figure~\ref{fig:ncmasses} shows similar plots for a non-cocoon bounded
simulation.  These figures are similar to Figure 4
of~\cite{2002MNRAS.332..271R} and primarily differ due to our
increased outer radius (and the plots are continued until later
times).  
The upper-left panel in each figure shows the total mass
of the cocoon material compared to the amount of injected material.
In both cases, this increases until the jet turns off and then begins
to decrease.  For the cocoon-bounded source, this decrease is mostly
steady for the entire lifetime of the simulation.  For the non-cocoon
source, the drop is much greater early on, so that about halfway through
the simulation most of the high entropy material has mixed with the
background and thus the decay rate has lessened.  This enhanced mixing
is a direct consequence of the lack of a well defined contact
discontinuity in the non-cocoon sources.  Note that the cocoon mass
does not correspond to the cocoon mass plot
in~\cite{2002MNRAS.332..271R} as we are interested in the total cocoon
mass, not the change in cocoon mass.

Also shown in Figs.~\ref{fig:cmasses} and \ref{fig:ncmasses}
(upper-right panel) is the length of the region containing jet
material.  In the cocoon bounded sources, the length of the cocoon
evolves in a smooth manner, gradually decelerating as the cocoon comes
into pressure balance with the ambient ICM.  The discontinuity at late
times occurs well into the phase in which the cocoon has transformed
into two buoyantly rising plumes and corresponds to the complete
mixing/dispersal of the leading part of the plume.  The evolution of
the length of the non-cocoon bounded sources, on the other hand,
suffered a sharp break at the time that the jet is shut off.

The two lower panels of Figs.~\ref{fig:cmasses} and \ref{fig:ncmasses}
show the change in energies for the jet material and the ICM material
respectively.  For the jet material, there is little difference in the
time-dependence of the energies for the two classes with one
exception.  In the non-cocoon case, the changes in internal and
kinetic energy peak at a comparable value although the kinetic drops
to almost zero very soon after.  For the cocoon case, the change in
kinetic energy never reaches a comparable level to the change in
internal energy.  Physically, this is close to the heart of the
difference between the cocoon and non-cocoon sources.  In each case, a
set amount of energy is injected by the jet (along with a set amount
of mass as seen in the potential change which is not very dramatic).
The evolution of the system then determines how this energy is split
between the internal energy and the kinetic energy (and eventually how
much is transfered to the ICM).  In the case of the cocoon, the hot
jet and shocked material is kept separate from the background, and
goes almost entirely into the internal energy of the shocked jet gas.  For
the non-cocoon case, the energy is split nearly evenly
between the internal and the kinetic energy.  This means there is no
longer enough internal energy available to inflate a cocoon of shocked
gas; more of the energy goes to the kinetic energy of the mostly bare
jet (with wispy areas of hot, shocked gas around it).  This is
consistent with the analytic estimate of~\citet{1992ApJ...392..458C}.

The time-dependence of the ICM energies show a greater difference
between the two cases.  The potential energy suffers a greater change
(within a few times the jet lifetime) in the cocoon case due to the
well-defined shell of ICM that is lifted up by the expanding cocoon,
while the non-cocoon case has a much more gradual change in the
potential energy.  The internal energy in the cocoon case drops back
down to the initial value around the same time the potential peaks
while the non-cocoon case has what looks like a slow trade-off between
the two.  Finally, the kinetic energy changes are comparable in the
two cases, as there appears to be a similar amount of disturbance in
the ICM regardless of the nature of the inflated structure.

\begin{figure*}
  \centering
  \epsscale{0.6}
  \plotone{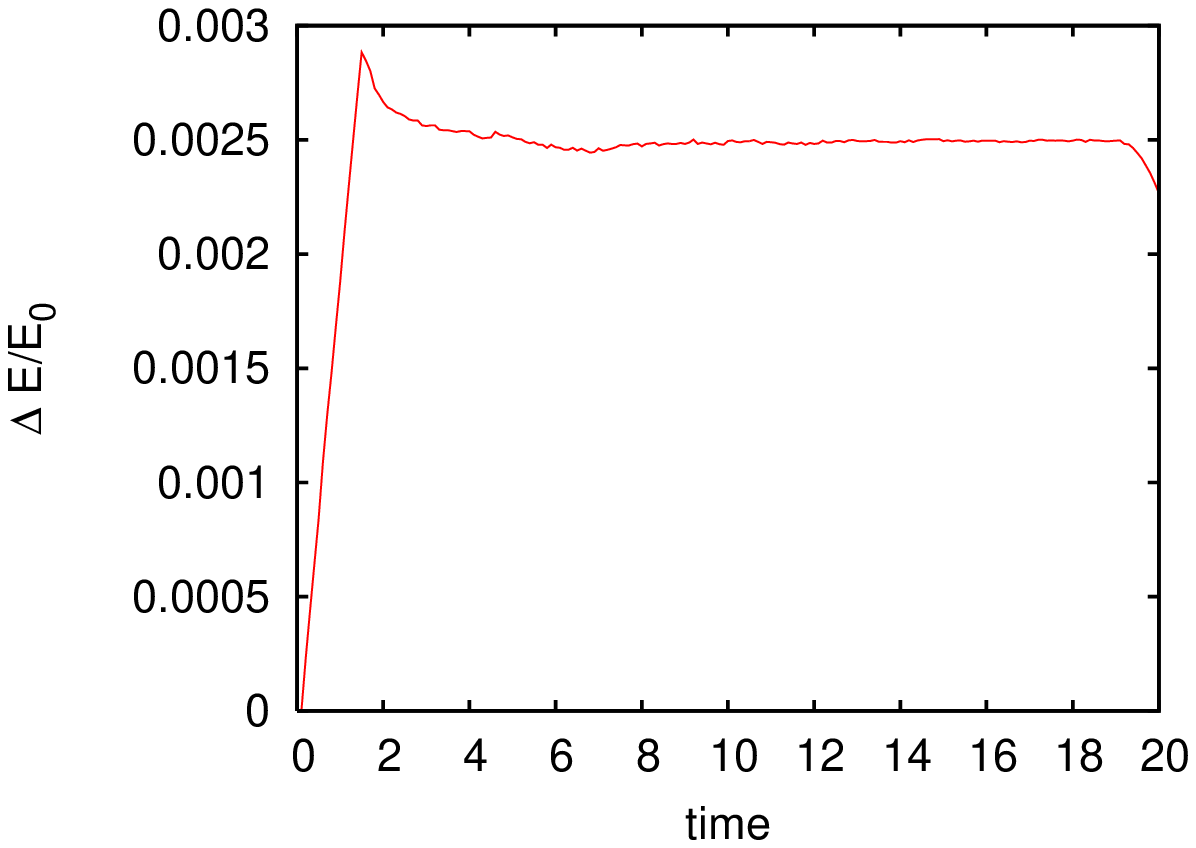}
  \caption{Fractional change in total energy within the simulated
  domain as a function of time.  All forms of energy (kinetic,
  internal and gravitational) for all of the gas in the computational
  domain are considered.  Shown here is the change in total energy
  compared with the initial time $\Delta E$ divided by the initial
  total energy $E_0$.}
  \label{fig:conserve}
\end{figure*}

In its standard configuration, ZEUS is not a strictly energy
conserving code.  We do not however believe that this is a major
impact on our results.  Fig.~\ref{fig:conserve} shows that during the
majority of time during the simulation, total energy is conserved to a
sufficient degree.  At very early times, the jet injects hot material
onto the grid which is clearly not conservative.  At very late times,
material (and sound waves carrying energy) leave the outer boundary of
the grid.  During  the time in between, the energy change remains flat
with only minor variations (at the $10^{-5}$ level), showing that
energy is mostly conserved.

\subsection{Entropy evolution and thermalization efficiencies}
\label{sec:en}

We now examine the effect of the jets on the entropy of the background
gas.  First, we define the specific entropy,
\begin{equation}
  \Delta S=S_1-S_0=\log{\sigma_1}-\log{\sigma_0},
\end{equation}
where $\sigma$ is the specific entropy index from
Equation~\ref{eq:ent}.

\begin{figure*}
  \centering
  \epsscale{0.6}
  \plotone{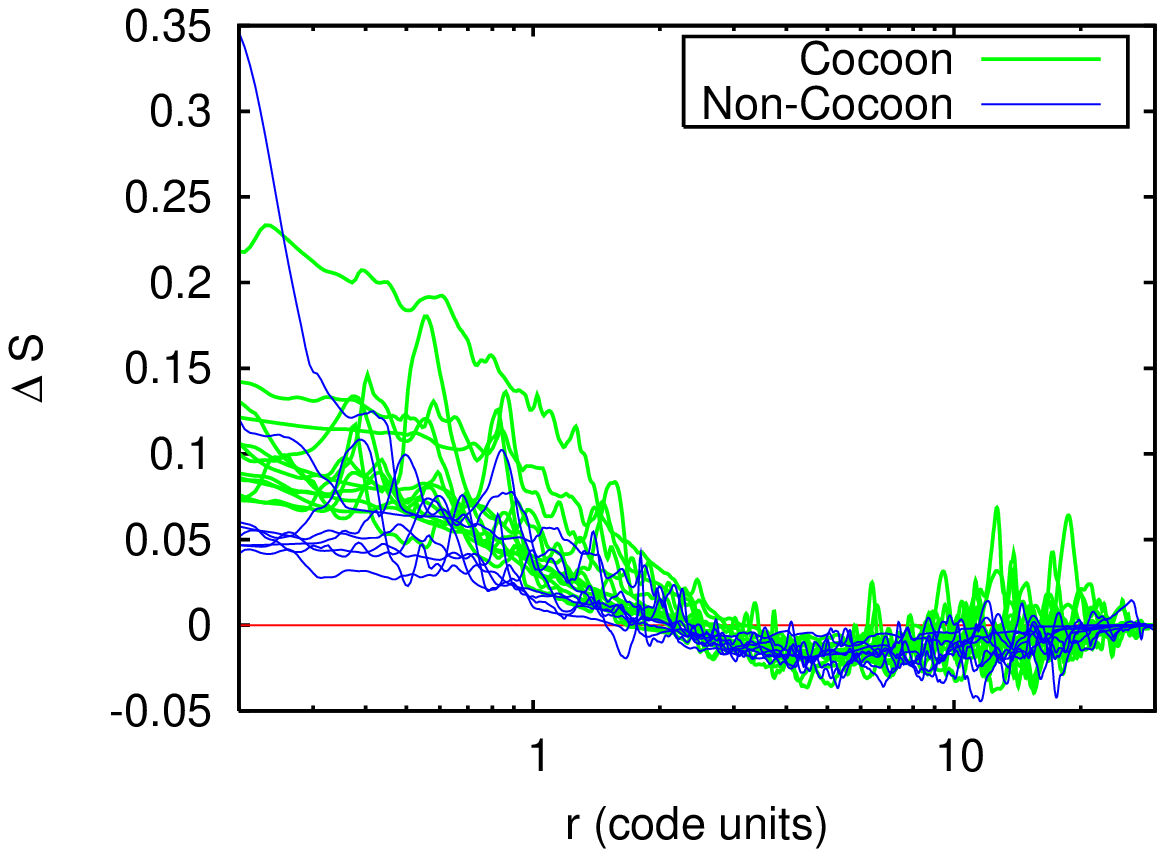}
  \caption{Angle averaged change in entropy.  See discussion in text
    for full description.}
  \label{fig:scomp}
\end{figure*}

Figure~\ref{fig:scomp} shows the entropy difference (compared to the
initial condition) at the final output of the simulations.  At this
point, the systems were allowed to evolve passively for nineteen times
the active lifetime of the jet.  For each simulation, the total
entropy at each radius was calculated and then averaged over the
angular coordinate.  The split between the different morphologies is
evident in this figure.  Within the core, the cocoon sources have a
higher $\Delta S$ overall.  Further discussion of this issue will be
reserved for Section~\ref{sec:disc}.

\begin{figure*}
  \centering
  \epsscale{0.6}
  \plotone{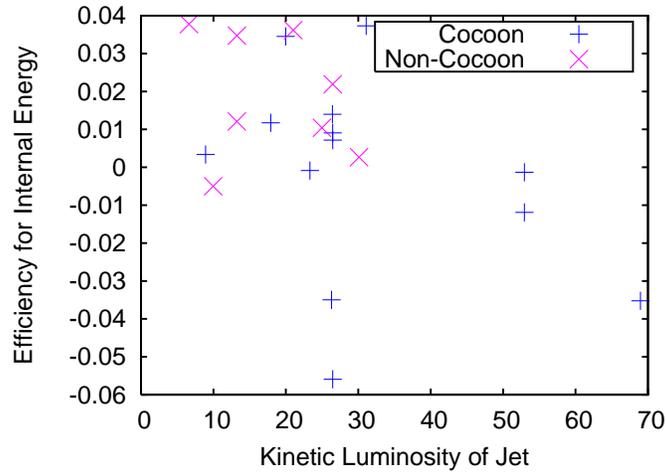}
  \caption{Efficiency for change in Internal Energy vs.\,\,Kinetic
  Luminosity for ambient medium.}
  \label{fig:change-int}
\end{figure*}

\begin{figure*}
  \centering
  \epsscale{0.6}
  \plotone{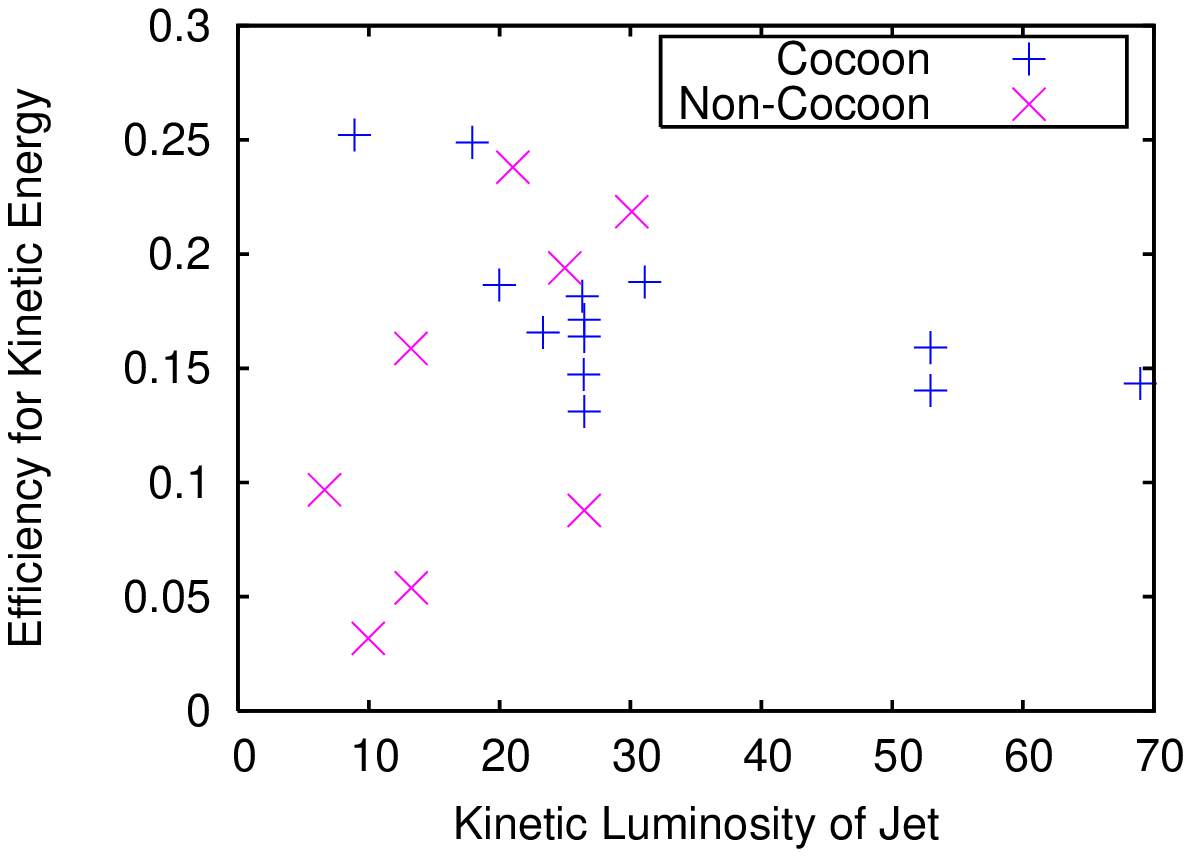}
  \caption{Efficiency for change in Kinetic Energy vs.\,\,Kinetic
  Luminosity for ambient medium.}
  \label{fig:change-kin}
\end{figure*}

\begin{figure*}
  \centering
  \epsscale{0.6}
  \plotone{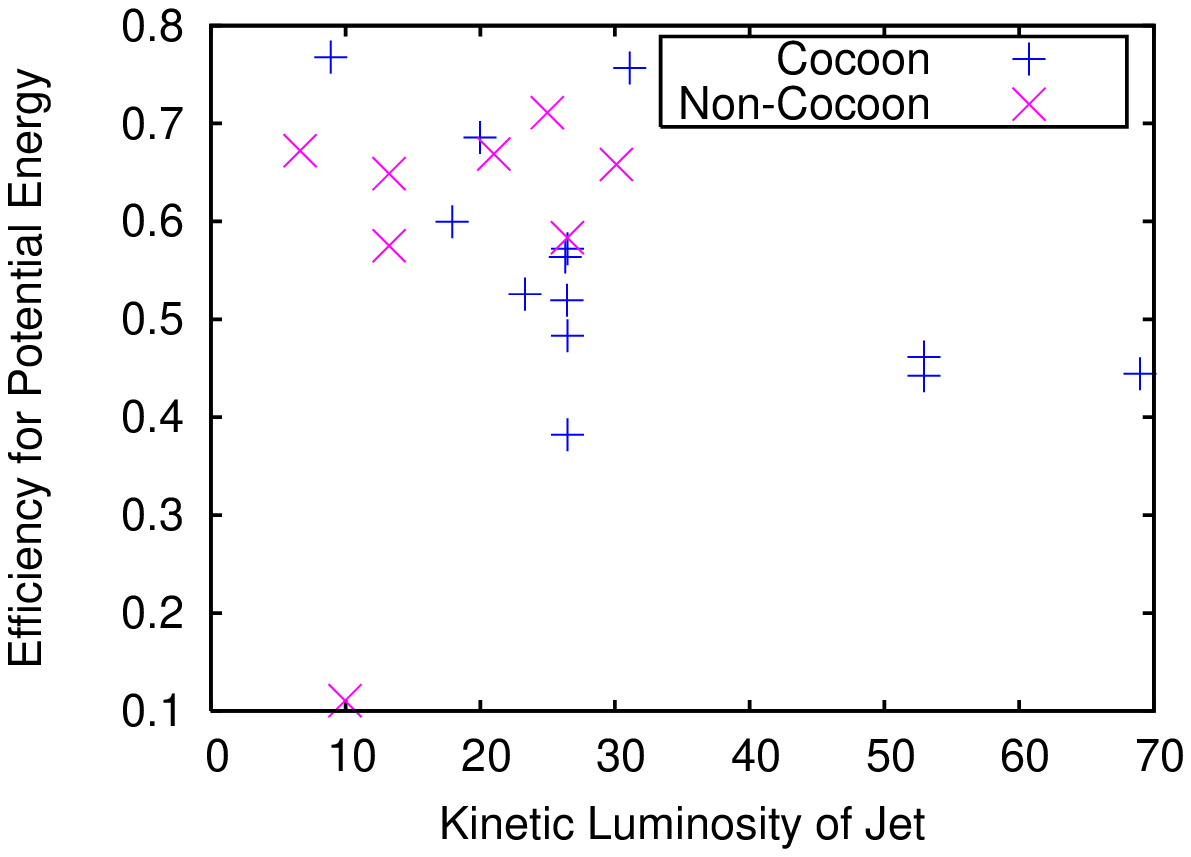}
  \caption{Efficiency for change in Potential Energy vs.\,\,Kinetic
  Luminosity for ambient medium.}
  \label{fig:change-pot}
\end{figure*}

As a final look at the effect of the jets on the cluster energetics,
we show the efficiencies for conversion of the injected energy into
various forms within the ICM as measured at the final time for
both the cocoon and non-cocoon sources.  This is defined as the change
in each type of energy for the ICM material divided by the total
amount of energy injected by the jet at the final output time of the
simulation.  Figures~\ref{fig:change-int}, \ref{fig:change-kin} and
\ref{fig:change-pot} show the efficiency for conversion of injected
energy into ICM internal energy, kinetic energy and potential energy,
respectively.  There is no clear segregation between the two types
although the cocoon sources do appear to cluster together somewhat.
In all cases, a large fraction of the injected energy (50--80\%) ends
as gravitational potential energy of the ICM as it expands in response
to the AGN heating.

\section{Discussion}
\label{sec:disc}

In the (resolved) simulations, we see a split in the morphology of the
jet-created structures.  As we saw in the parameter space plot
(Figure~\ref{fig:pspace}), the presence or absence of a cocoon is not
related to the jet speed or power in a simple matter.  So the forms
that we see do not tell us directly about the parameters of the jet
that formed them.  What they do tell us is something about the way the
jet interacted with the ICM.

The cocoon itself is formed by shocks.  The supersonic jet initially
flows freely through the ICM.  After traveling some distance (usually
short on the scale of these simulations), the jet will build up enough
material in front of it that it can no longer flow freely.  The
material at the head of the jet is shock heated and expands.  Up to
this point, the evolution is the same for both cocoon and non-cocoon
sources.  As this shocked material expands, the evolution diverges.

The hot, shocked material expands into the surrounding medium.  At the
boundaries of the shocked material, fluid instabilities (i.e., the
Rayleigh-Taylor (RT) and Kelvin-Helmholtz (KH) instabilities) work to
shred the contact discontinuity and destroy the forming bubble.  In
the cocoon bounded sources, the bubble is inflated before it is
shredded and continues to inflate until it reaches pressure
equilibrium with the background.  While active, the jet deposits
energy only at point where it comes into contact with the shocked ICM
and freshly shocked material flows back from there, continuing to
inflate the bubble.  Once the jet ceases, the bubble floats buoyantly
from the cluster center.

In the case of non-cocoon sources, the jet is unable to inflate a
bubble.  If the jet cannot deposit enough energy at the contact point
to inflate the bubble, the fluid instabilities will mix the shocked
material with the background material, preventing the bubble from
ever growing.  This leaves areas of hot, shocked gas that can rise,
but never produces a single coherent structure.

The clearest distinction in the thermodynamics of cocoon and
non-cocoon sources can be seen in the spatial distribution of the
entropy enhancement.  We find that in the core regions, cocoon sources
cause a greater enhancement of the ICM entropy than non-cocoon sources
although there is some overlap.  Outside the core (around $r=5$), both
appear the same and cause a slight negative change in the entropy.
This negative change corresponds to lower entropy material that has
been pushed upwards by the expansion of the core.  For the non-cocoon
sources, this change persists to the edge of the simulation.  Some,
but not all, of the cocoon sources have a spiky, positive change when
approaching the edge of the grid --- this corresponds to the remnants
of the buoyant cocoon that, in some cases, remains intact until the
end of the simulation.  It is important to note that the comparison of
overall entropy change was done at the final output ($t=20$).  This is
long after our classifications were done.  At this point, mixing has
disturbed the structure enough that it would no longer be possible to
separate the simulations into cocoon or non-cocoon categories.  The
fact that we still see a difference at this late a time shows that the
jets have long lasting effects on the energetics of the ICM and that
these effects vary based on the jet properties in a way that is more
complicated than simply depending on jet power.

The energy efficiencies (Figures~\ref{fig:change-int},
\ref{fig:change-kin}, and \ref{fig:change-pot}) are initially
puzzling.  The non-cocoon simulations generally seem to coincide with
the highest efficiency cocoon simulations for internal and potential
energy efficiency.  For the kinetic energy
(Figure~\ref{fig:change-kin}), the non-cocoon runs seem to have both
close to the highest values and the lowest values.  This is in
contrast with the situation for entropy where the cocoon simulations
have a greater impact.  It is fairly clear why the cocoon sources have
a great entropy enhancement overall: shocks.  The cocoon is formed by
shocked gas, so jets that are capable of forming cocoons are also
generally the ones that shock more ICM (although this is not the
entire story as the rate of shocked gas vs. mixed gas matters as
well).  It is not as immediately obvious why jets that are good at
increasing the entropy are not always as good at increasing the energy
and why there can be a spread in efficiencies for jets with similar
parameters.  To see why, we must go back to Equation~\ref{eq:ent}, the
definition of the specific entropy index (keeping in mind that the
pressure, $P$, is directly proportional to the internal energy.  For a
given change in internal energy, the change in entropy can be higher
(or lower) by having it take place at a lower (or higher) density.  So
if the shocks happen higher in the potential well (and hence at a
lower density), it is possible to have a greater change in entropy for
the same amount of energy input.  The spread in kinetic energies shows
that it is possible to stir up the gas significantly with both the
cocoon and the non-cocoon producing jets.  Also, as stated in the
discussion, the cocoon producing jets leave more energy available as
internal energy which can go towards shocks while the non-cocoon
producing jets have more of their energy in kinetic energy which does
not go to shocks and therefore is not available to increase the
entropy.  It is also the case that different jets will cause a
different amount of material and energy to be lost from the system at
the outer edge of the grid.  We consider it to be reasonable to count
this as lost energy as anything that has not managed to heat the gas
by the time it reaches the outer edge is clearly not going to be able
to change the temperature of the system.

This disconnect between entropy increase and energy increase also fits
in with the results from~\cite{2006ApJ...645...83V}, which showed that
even with large amounts of energy available, a jet can produce large,
well formed cocoons yet fail to significantly heat the ICM.  The large
cocoons indicate a large change in entropy, but the the central
regions were still not heated significantly.  This helps to reaffirm
the point that in pure hydrodynamics, it is hard to energetically
couple powerful jets to the {\it core} regions of the ICM (where it is
needed to solve the cooling flow problem!).

Along with the energetic issues, we may also compare these simulations
to observations in a broad sense.  Classically, radio galaxies are
split into two categories: Faranoff-Riley Type I (FR I) and
Faranoff-Riley Type II (FR II)~\citep{1974MNRAS.167P..31F}.  Lacking
any information on luminosities and radio emission, we cannot directly
compare our simulations to these classes of real radio galaxies, but
we can note some interesting correlations.  FR II galaxies encompass
the large, back-to-back classical double sided sources.  Our
cocoon-bounded sources are reminiscent of the classical doubles.  The
less lobe dominated and often bent FR I sources share a similar
structure with our non-cocoon sources.  Although our classifications
were done based on the passive phase of the system, we can also
compare the jets from the active phase to the FR I/FR II divide.  
As seen in Fig.~\ref{fig:pspace} and~\ref{fig:pspace2}, slow, heavy,
and lower luminosity jets (the non-cocoon cases) should correspond to
the FR I sources while fast, light, and more luminous jets should
correspond to the FR II type.  As FR II galaxies have higher
luminosity jets~\citep{1996AJ....112....9L}, this is at least
consistent with our simulations.
Even without a definite
correspondence between the two, we do show that it is possible to
produce qualitatively different structures with differences in the jet
parameters without any environmental changes.

Two additional simulations (run 8 and run 9) were performed where all
the parameters were kept the same as our standard case (run 1), but
the active time of the jet was varied.  In run 8, the jet had an
active time of only $t_j=0.5$ while run 9 had an active time of
$t_j=1.5$.  In both cases, these simulations fall in the same
cocoon-bounded category as run 1.  This give us some confidence that
our results are not simply based on the amount of
total energy inject by the jet, but instead depend on the detailed
hydrodynamic interactions between the jets and the ambient material.

\section{Conclusion}
\label{sec:conc}

We have performed a large number of axisymmetric simulations of AGN
jets in cluster atmosphere.  By limiting ourselves to pure
hydrodynamics with no radiative cooling, we are able to carefully
study the interaction of the atmosphere with the jets and the
formation of cocoons and bubbles in the cluster.  With the continued
interest in the probable role of jets and bubbles in the cooling flow
problem, it is important that we understand their dynamics as clearly
as possible.

Our main results are summarized in Figures~\ref{fig:pspace} and
\ref{fig:scomp}.  The morphology vs.\,\,jet parameters plot shows that we
are able to produce two distinct classes of structures, cocoon bound
and non-cocoon bound, by varying the initial jet parameters.  This
distinction does not change with a change in total injected energy (as
seen when we vary the jet lifetime) and is a function of both jet
parameters (Mach number and kinetic luminosity).  We can also visually
draw comparisons between our two classes and the FR I and FR II
distinction amongst real radio galaxies.  The angle averaged entropy
change vs. radius shows that along with the split in morphology, the
effect on the energetics of the ICM also depends on the jet
parameters.  We also see that even short lived single burst jets are
capable of a long-lived enhancement of the entropy of a cluster core.

\acknowledgements
We would like to thank the developers of ZEUS-3D and NCSA for
providing the code we used.  We would like to thank the anonymous
referee for useful comments that allowed us to enhance the discussion
in several ways.  All simulation in this paper were
performed on the Beowulf cluster (``the Borg'') in the Department of
Astronomy, University of Maryland.  We thank support from Chandra
Theory and Modeling Program under grants TM4-5007X and TM7-8009X.

\bibliographystyle{apj}
\bibliography{paper2d}

\begin{thebibliography}{40}
\expandafter\ifx\csname natexlab\endcsname\relax\def\natexlab#1{#1}\fi

\bibitem[{{Basson} \& {Alexander}(2003)}]{2003MNRAS.339..353B}
{Basson}, J.~F. \& {Alexander}, P. 2003, \mnras, 339, 353

\bibitem[{{Begelman} \& {Cioffi}(1989)}]{1989ApJ...345L..21B}
{Begelman}, M.~C. \& {Cioffi}, D.~F. 1989, \apjl, 345, L21

\bibitem[{{Benson} {et~al.}(2003){Benson}, {Bower}, {Frenk}, {Lacey}, {Baugh},
  \& {Cole}}]{2003ApJ...599...38B}
{Benson}, A.~J., {Bower}, R.~G., {Frenk}, C.~S., {Lacey}, C.~G., {Baugh},
  C.~M., \& {Cole}, S. 2003, \apj, 599, 38

\bibitem[{{Blanton} {et~al.}(2001){Blanton}, {Sarazin}, {McNamara}, \&
  {Wise}}]{2001ApJ...558L..15B}
{Blanton}, E.~L., {Sarazin}, C.~L., {McNamara}, B.~R., \& {Wise}, M.~W. 2001,
  \apjl, 558, L15

\bibitem[{{Br{\"u}ggen} \& {Kaiser}(2001)}]{2001MNRAS.325..676B}
{Br{\"u}ggen}, M. \& {Kaiser}, C.~R. 2001, \mnras, 325, 676

\bibitem[{{Br{\"u}ggen} \& {Kaiser}(2002)}]{2002Natur.418..301B}
---. 2002, \nat, 418, 301

\bibitem[{{Carvalho} \& {O'Dea}(2002)}]{2002ApJS..141..371C}
{Carvalho}, J.~C. \& {O'Dea}, C.~P. 2002, \apjs, 141, 371

\bibitem[{{Choi} {et~al.}(2004){Choi}, {Reynolds}, {Heinz}, {Rosenberg},
  {Perlman}, \& {Yang}}]{2004ApJ...606..185C}
{Choi}, Y., {Reynolds}, C.~S., {Heinz}, S., {Rosenberg}, J.~L., {Perlman},
  E.~S., \& {Yang}, J. 2004, \apj, 606, 185

\bibitem[{{Churazov} {et~al.}(2001){Churazov}, {Br{\"u}ggen}, {Kaiser},
  {B{\"o}hringer}, \& {Forman}}]{2001ApJ...554..261C}
{Churazov}, E., {Br{\"u}ggen}, M., {Kaiser}, C.~R., {B{\"o}hringer}, H., \&
  {Forman}, W. 2001, \apj, 554, 261

\bibitem[{{Cioffi} \& {Blondin}(1992)}]{1992ApJ...392..458C}
{Cioffi}, D.~F. \& {Blondin}, J.~M. 1992, \apj, 392, 458

\bibitem[{{Clarke} {et~al.}(1997){Clarke}, {Harris}, \&
  {Carilli}}]{1997MNRAS.284..981C}
{Clarke}, D.~A., {Harris}, D.~E., \& {Carilli}, C.~L. 1997, \mnras, 284, 981

\bibitem[{{Cowie} \& {Binney}(1977)}]{1977ApJ...215..723C}
{Cowie}, L.~L. \& {Binney}, J. 1977, \apj, 215, 723

\bibitem[{{Dalla Vecchia} {et~al.}(2004){Dalla Vecchia}, {Bower}, {Theuns},
  {Balogh}, {Mazzotta}, \& {Frenk}}]{2004MNRAS.355..995D}
{Dalla Vecchia}, C., {Bower}, R.~G., {Theuns}, T., {Balogh}, M.~L., {Mazzotta},
  P., \& {Frenk}, C.~S. 2004, \mnras, 355, 995

\bibitem[{{Fabian}(1994)}]{1994ARA&A..32..277F}
{Fabian}, A.~C. 1994, \araa, 32, 277

\bibitem[{{Fabian} \& {Nulsen}(1977)}]{1977MNRAS.180..479F}
{Fabian}, A.~C. \& {Nulsen}, P.~E.~J. 1977, \mnras, 180, 479

\bibitem[{{Fabian} {et~al.}(2003){Fabian}, {Sanders}, {Allen}, {Crawford},
  {Iwasawa}, {Johnstone}, {Schmidt}, \& {Taylor}}]{2003MNRAS.344L..43F}
{Fabian}, A.~C., {Sanders}, J.~S., {Allen}, S.~W., {Crawford}, C.~S.,
  {Iwasawa}, K., {Johnstone}, R.~M., {Schmidt}, R.~W., \& {Taylor}, G.~B. 2003,
  \mnras, 344, L43

\bibitem[{{Fabian} {et~al.}(2000){Fabian}, {Sanders}, {Ettori}, {Taylor},
  {Allen}, {Crawford}, {Iwasawa}, {Johnstone}, \& {Ogle}}]{2000MNRAS.318L..65F}
{Fabian}, A.~C., {Sanders}, J.~S., {Ettori}, S., {Taylor}, G.~B., {Allen},
  S.~W., {Crawford}, C.~S., {Iwasawa}, K., {Johnstone}, R.~M., \& {Ogle}, P.~M.
  2000, \mnras, 318, L65

\bibitem[{{Fabian} {et~al.}(2005){Fabian}, {Sanders}, {Taylor}, \&
  {Allen}}]{2005MNRAS.360L..20F}
{Fabian}, A.~C., {Sanders}, J.~S., {Taylor}, G.~B., \& {Allen}, S.~W. 2005,
  \mnras, 360, L20

\bibitem[{{Fanaroff} \& {Riley}(1974)}]{1974MNRAS.167P..31F}
{Fanaroff}, B.~L. \& {Riley}, J.~M. 1974, \mnras, 167, 31P

\bibitem[{{Heinz} {et~al.}(2006){Heinz}, {Br{\"u}ggen}, {Young}, \&
  {Levesque}}]{2006MNRAS.373L..65H}
{Heinz}, S., {Br{\"u}ggen}, M., {Young}, A., \& {Levesque}, E. 2006, \mnras,
  373, L65

\bibitem[{{Heinz} {et~al.}(2002){Heinz}, {Choi}, {Reynolds}, \&
  {Begelman}}]{2002ApJ...569L..79H}
{Heinz}, S., {Choi}, Y., {Reynolds}, C.~S., \& {Begelman}, M.~C. 2002, \apjl,
  569, L79

\bibitem[{{Hicks} \& {Mushotzky}(2005)}]{2005ApJ...635L...9H}
{Hicks}, A.~K. \& {Mushotzky}, R. 2005, \apjl, 635, L9

\bibitem[{{Jones} \& {De Young}(2005)}]{2005ApJ...624..586J}
{Jones}, T.~W. \& {De Young}, D.~S. 2005, \apj, 624, 586

\bibitem[{{Ledlow} \& {Owen}(1996)}]{1996AJ....112....9L}
{Ledlow}, M.~J. \& {Owen}, F.~N. 1996, \aj, 112, 9

\bibitem[{{McNamara} {et~al.}(2000){McNamara}, {Wise}, {Nulsen}, {David},
  {Sarazin}, {Bautz}, {Markevitch}, {Vikhlinin}, {Forman}, {Jones}, \&
  {Harris}}]{2000ApJ...534L.135M}
{McNamara}, B.~R., {Wise}, M., {Nulsen}, P.~E.~J., {David}, L.~P., {Sarazin},
  C.~L., {Bautz}, M., {Markevitch}, M., {Vikhlinin}, A., {Forman}, W.~R.,
  {Jones}, C., \& {Harris}, D.~E. 2000, \apjl, 534, L135

\bibitem[{{McNamara} {et~al.}(2001){McNamara}, {Wise}, {Nulsen}, {David},
  {Carilli}, {Sarazin}, {O'Dea}, {Houck}, {Donahue}, {Baum}, {Voit},
  {O'Connell}, \& {Koekemoer}}]{2001ApJ...562L.149M}
{McNamara}, B.~R., {Wise}, M.~W., {Nulsen}, P.~E.~J., {David}, L.~P.,
  {Carilli}, C.~L., {Sarazin}, C.~L., {O'Dea}, C.~P., {Houck}, J., {Donahue},
  M., {Baum}, S., {Voit}, M., {O'Connell}, R.~W., \& {Koekemoer}, A. 2001,
  \apjl, 562, L149

\bibitem[{{Mizuta} {et~al.}(2001){Mizuta}, {Yamada}, \&
  {Takabe}}]{2001JKAS...34..329M}
{Mizuta}, A., {Yamada}, S., \& {Takabe}, H. 2001, Journal of Korean
  Astronomical Society, 34, 329

\bibitem[{{O'Dea} {et~al.}(2004){O'Dea}, {Baum}, {Mack}, {Koekemoer}, \&
  {Laor}}]{2004ApJ...612..131O}
{O'Dea}, C.~P., {Baum}, S.~A., {Mack}, J., {Koekemoer}, A.~M., \& {Laor}, A.
  2004, \apj, 612, 131

\bibitem[{{Omma} \& {Binney}(2004)}]{2004MNRAS.350L..13O}
{Omma}, H. \& {Binney}, J. 2004, \mnras, 350, L13

\bibitem[{{Omma} {et~al.}(2004){Omma}, {Binney}, {Bryan}, \&
  {Slyz}}]{2004MNRAS.348.1105O}
{Omma}, H., {Binney}, J., {Bryan}, G., \& {Slyz}, A. 2004, \mnras, 348, 1105

\bibitem[{{Reynolds} {et~al.}(2002){Reynolds}, {Heinz}, \&
  {Begelman}}]{2002MNRAS.332..271R}
{Reynolds}, C.~S., {Heinz}, S., \& {Begelman}, M.~C. 2002, \mnras, 332, 271

\bibitem[{{Reynolds} {et~al.}(2005){Reynolds}, {McKernan}, {Fabian}, {Stone},
  \& {Vernaleo}}]{2005MNRAS.357..242R}
{Reynolds}, C.~S., {McKernan}, B., {Fabian}, A.~C., {Stone}, J.~M., \&
  {Vernaleo}, J.~C. 2005, \mnras, 357, 242

\bibitem[{{Robinson} {et~al.}(2004){Robinson}, {Dursi}, {Ricker}, {Rosner},
  {Calder}, {Zingale}, {Truran}, {Linde}, {Caceres}, {Fryxell}, {Olson},
  {Riley}, {Siegel}, \& {Vladimirova}}]{2004ApJ...601..621R}
{Robinson}, K., {Dursi}, L.~J., {Ricker}, P.~M., {Rosner}, R., {Calder}, A.~C.,
  {Zingale}, M., {Truran}, J.~W., {Linde}, T., {Caceres}, A., {Fryxell}, B.,
  {Olson}, K., {Riley}, K., {Siegel}, A., \& {Vladimirova}, N. 2004, \apj, 601,
  621

\bibitem[{{Ruszkowski} {et~al.}(2004){Ruszkowski}, {Br{\" u}ggen}, \&
  {Begelman}}]{2004ApJ...611..158R}
{Ruszkowski}, M., {Br{\" u}ggen}, M., \& {Begelman}, M.~C. 2004, \apj, 611, 158

\bibitem[{{Sternberg} {et~al.}(2007){Sternberg}, {Pizzolato}, \&
  {Soker}}]{2007ApJ...656L...5S}
{Sternberg}, A., {Pizzolato}, F., \& {Soker}, N. 2007, \apjl, 656, L5

\bibitem[{{Stone} \& {Norman}(1992{\natexlab{a}})}]{1992ApJS...80..753S}
{Stone}, J.~M. \& {Norman}, M.~L. 1992{\natexlab{a}}, \apjs, 80, 753

\bibitem[{{Stone} \& {Norman}(1992{\natexlab{b}})}]{1992ApJS...80..791S}
---. 1992{\natexlab{b}}, \apjs, 80, 791

\bibitem[{{Vernaleo} \& {Reynolds}(2006)}]{2006ApJ...645...83V}
{Vernaleo}, J.~C. \& {Reynolds}, C.~S. 2006, \apj, 645, 83

\bibitem[{{Young} {et~al.}(2002){Young}, {Wilson}, \&
  {Mundell}}]{2002ApJ...579..560Y}
{Young}, A.~J., {Wilson}, A.~S., \& {Mundell}, C.~G. 2002, \apj, 579, 560

\bibitem[{{Zanni} {et~al.}(2005){Zanni}, {Murante}, {Bodo}, {Massaglia},
  {Rossi}, \& {Ferrari}}]{2005A&A...429..399Z}
{Zanni}, C., {Murante}, G., {Bodo}, G., {Massaglia}, S., {Rossi}, P., \&
  {Ferrari}, A. 2005, \aap, 429, 399

\end{thebibliography}

\end{document}